\documentclass[aps,prd,preprint,nofootinbib]{revtex4-1}
\pdfoutput=1
\usepackage{graphicx}
\usepackage{psfrag}
\usepackage{float}
\usepackage{subfigure}
\usepackage{array}
\usepackage{graphicx}
\usepackage{amsmath}
\usepackage{amssymb}
\usepackage{mathrsfs}
\usepackage{bm}
\usepackage{slashed}
\usepackage{threeparttable}
\usepackage{multirow}
\usepackage[colorlinks, linkcolor=black, anchorcolor=black, citecolor=black]{hyperref}
\usepackage[amsmath,thmmarks,hyperref]{ntheorem}
\usepackage{soul}
\allowdisplaybreaks[4]

\begin{document}

\title{Bottomed baryon decays with invisible Majorana fermions}
\author{Geng Li\footnote[1]{ligeng@ucas.ac.cn}, Chia-Wei Liu, Chao-Qiang Geng\\}
\affiliation{School of Fundamental Physics and Mathematical Sciences, Hangzhou Institute for Advanced Study, UCAS, 310024 Hangzhou, China\\
	University of Chinese Academy of Sciences, 100190 Beijing, China
	\vspace{0.6cm}} 

\baselineskip=20pt

\begin{abstract}
We study the invisible Majorana fermions of $\chi$ in bottomed baryon decays with flavor-changing neutral currents (FCNCs) based on the model-independent effective Lagrangian between the quarks and invisible particles. From the bounds of the coupling constants extracted from the experiments, we examine the decay branching ratios of $\Lambda_b\to\Lambda\chi\chi$, $\Xi_b^{0(-)}\to\Xi^{0(-)}\chi\chi$, $\Lambda_b\to n \chi\chi$, $\Xi_b^{-}\to \Sigma^{-} \chi\chi$, $\Xi_b^{0}\to \Sigma^{0} \chi\chi$, and $\Xi_b^{0}\to \Lambda \chi\chi$, which can be as large as $6.3,~9.2,~5.7,~5.8,~2.7$, and $1.0\times10^{-5}$ for $m_\chi=2$ GeV, respectively. Some of these decays are accessible to the future experimental searches, such as Belle II.

\end{abstract}

\maketitle

%%%%%%%%%%%%%%%%%%%%%%%%%%%%%%%%%%
\section{Introduction}
%%%%%%%%%%%%%%%%%%%%%%%%%%%%%%%%%%
It is known that flavor-changing neutral current (FCNC) processes of long-lived particles would provide a window to observe new physics (NP) beyond the standard model (SM). These particles, such as ground-state mesons of $K,~B,~D$, and $B_c$, and baryons of $\Lambda_{c,b}$ and $\Xi_{c,b}$, decay through weak interactions, resulting in longer lifetimes and narrower decay widths. These FCNC decays may benefit the detections of NP. Hadronic FCNC decays include $c\to u$, $s\to d$, $b\to d$, and $b\to s$ processes at quark level. In the SM, dilepton FCNC modes have been widely studied theoretically~\cite{Buchalla:1993wq,Misiak:1999yg,Mott:2011cx,Khodjamirian:2012rm,Bordone:2016gaq} and experimentally~\cite{CLEO:2001kln,LHCb:2013hxr,LHCb:2015svh,LHCb:2018rym,BELLE:2019xld}. However, the neutrino ($\nu$) and anti-neutrino ($\bar \nu$) in the final states of the decays cannot be directly detected, but are treated as missing energy ($\slashed E$) in experiments. So far most experiments can only obtain the upper limits on the decay branching ratios associated with $\nu\bar\nu$~\cite{KOTO:2018dsc, NA62:2022zos, E949:2008btt, CLEO:2000yzg, BaBar:2004xlo, BaBar:2008wiw, BaBar:2010oqg, BaBar:2013npw, Belle:2007vmd, Belle:2013tnz,Belle:2017oht}. 

The experimental searches have given the strictest constraints on kaon FCNC decays. Recently, the upper bound on $K_L\to\pi^0\bar\nu\nu$ from the KOTO experiment at J-PARC~\cite{KOTO:2018dsc} has been given to be $\mathcal {B}(K_L\to\pi^0\bar\nu\nu)_{\rm KOTO}<3.0\times 10^{-9}$ at $90\%$ confidence level (C.L.), which is slightly greater than the SM prediction of $\mathcal {B}(K_L\to\pi^0\bar\nu\nu)_{\rm SM} = (3.4 \pm 0.6)\times10^{-11}$~\cite{Buras:2015qea}. On the other hand, the decay of $K^+\to \pi^+\bar\nu\nu$ has been measured, namely, $\mathcal {B}(K^+\to\pi^+\bar\nu\nu)_{\rm NA62}=(11.0^{+4.0}_{-3.5}({\rm stat})\pm 0.3(\rm syst))\times 10^{-11}$ at $68\%$ C.L. from the NA62 experiment at CERN~\cite{NA62:2022zos} and $\mathcal {B}(K^+\to\pi^+\bar\nu\nu)_{\rm E949}= (17.3^{+11.5}_{-10.5})\times 10^{-11}$ from the E949 experiment at BNL~\cite{E949:2008btt}. These results are consistent with the SM prediction of $\mathcal {B}(K^+\to\pi^+\bar\nu\nu)_{\rm SM}= (8.4\pm1.0)\times 10^{-11}$~\cite{Buras:2015qea} within one standard deviation. It is clear that the room for NP in $K \to \pi\slashed E $ has become quite small. 

However, the searches for NP in the FCNC decay processes of charmed and bottomed hadrons would still be possible. For example, the charmed meson and hyperon decays associated with $\slashed E$ have been analyzed in Ref.~\cite{Faisel:2020php}. The invisible decays of bottomed mesons have attracted more attention experimentally. For example, the upper bounds of branching ratios of $B\to K^{(*)}, \pi, \rho$ modes have been given by the CLEO~\cite{CLEO:2000yzg}, BarBar~\cite{BaBar:2004xlo, BaBar:2008wiw, BaBar:2010oqg, BaBar:2013npw}, Belle~\cite{Belle:2007vmd, Belle:2013tnz, Belle:2017oht} and Belle II~\cite{Belle-II:2021rof} collaborations. Particularly, the Belle II~\cite{Belle-II:2018jsg} collaboration has estimated that the sensitivity for the measurement of the branching ratios of $B^{0(+)} \to K^{(*)0(+)}\bar\nu\nu$ processes can be increased by $25-30\%$ in the near future, when assuming that $5~\rm ab^{-1}$ of data will be taken on the $\Upsilon (5S)$ resonance. In additional, the future $e^+ e^-$ colliders, such as the FCC-ee~\cite{FCC:2018byv, FCC:2018evy, Blondel:2014bra} experiment, have shown the ability of precise measurements of FCNC processes. The current measurements of the experimental bounds which are listed as the first column in Table~\ref{exp} are cited from Refs.~\cite{BaBar:2013npw, Belle:2017oht, Belle:2013tnz}. The SM predictions cited from~\cite{Bause:2021cna, Du:2015tda} contains both short-distance and long-distance contributions. 
\begin{table}[h]
	\setlength{\tabcolsep}{0.5cm}
	\caption{The branching ratios ($\mathcal B$) (in units of $10^{-6}$) of $B$ decays involving missing energy.}
	\centering
	\begin{tabular*}{\textwidth}{@{}@{\extracolsep{\fill}}ccc}
		\hline\hline
		Experimental bound~\cite{BaBar:2013npw,Belle:2017oht,Belle:2013tnz}\footnote{These experimental bounds are adopted by PDG-live, which are not certainly the latest or strictest constraints.}&SM prediction~\cite{Bause:2021cna, Du:2015tda}&Invisible particles bound\\
		\hline
		$\mathcal {B}(B^\pm\to K^\pm\slashed E)<16$& $\mathcal {B}(B^\pm\to K^\pm\nu \bar{\nu})=4.73 \pm 0.56$  & $\mathcal {B}(B^\pm\to K^\pm\chi\chi)<11.8$	\\
		$\mathcal {B}(B^\pm\to \pi^\pm\slashed E)<14$& $\mathcal {B}(B^\pm\to \pi^\pm\nu \bar{\nu})=8.12 \pm 0.01$  & $\mathcal {B}(B^\pm\to \pi^\pm\chi\chi)<5.89$	\\
		$\mathcal {B}(B^\pm\to K^{*\pm}\slashed E)<40$& $\mathcal {B}(B^\pm\to K^{*\pm}\nu \bar{\nu})=8.93 \pm 1.07 $  & $\mathcal {B}(B^\pm\to K^{*\pm}\chi\chi)<32.1$	\\
		$\mathcal {B}(B^\pm\to \rho^\pm\slashed E)<30$&$\mathcal {B}(B^\pm\to \rho^\pm \nu \bar{\nu})=0.48\pm 0.18$  & $\mathcal {B} (B^\pm\to \rho^\pm \chi\chi)<29.7$	\\
		\hline\hline
		\label{exp}
	\end{tabular*}
\end{table}
In Table~\ref{exp}, the differences between the first and second columns indicate that there are some rooms for new invisible particles of $\chi$ (shown as the third column) emitted in such processes. In Refs.~\cite{Bird:2004ts, Bird:2006jd, Badin:2010uh, Kamenik:2011vy, Gninenko:2015mea, Bertuzzo:2017lwt, Barducci:2018rlx, Li:2018hgu, Li:2020dpc, Li:2021sqe}, the effects of the invisible particles with various spins in the FCNC and neutral meson annihilation processes have been explored. While most of these previous studies in the literature are focused on mesons. There have been no related researches on bottomed baryons up to now. In this paper, we generalize the experimental upper bounds from $B$ mesons to the corresponding decay modes of bottomed baryons, namely, $\Lambda_b$ and $\Xi_b$. These modes are accessible for the Belle II collaboration~\cite{Belle-II:2018jsg}, which will be able to obtain more sensitive results in future projects. Clearly, in the near future the experiments on bottomed baryons would provide an interesting window to probe with invisible particles.

In this work, we consider the bottomed baryonic FCNC decays of ${\bf B}_b\to\mathcal {\bf B}_n\chi\chi$, where ${\bf B}_{n(b)}$ are (bottomed) baryons and $\chi$ represent light invisible particles, which are assumed to be Majorana fermions. Phenomenologically, these new invisible fermions of $\chi$ can weakly interact with the SM fermions via a mediator, which can be a scalar~\cite{Matsumoto:2018acr}, pseudoscalar~\cite{Yang:2016wrl}, vector or axial-vector~\cite{Chala:2015ama} particle. In our study, we will concentrate on a general model-independent approach to introduce the effective Lagrangian, which contains all possible currents involving the invisible fermions with the coupling constants extracted from the experiments. 

The paper is organized as follows: In Sec. II, we obtain the SM expectations of ${\bf B}_b\to\mathcal {\bf B}_n\bar\nu\nu$. In Sec. III, we first construct the effective Lagrangian, which describes the coupling between the quarks and light invisible fermions. We then present the numerical results of the upper limits for the decay branching ratios of ${\bf B}_b\to\mathcal {\bf B}_n\chi\chi$. The hadronic transition matrix elements are evaluated based on the QCD light-cone sum rules (LCSR) and modified bag model (MBM). Finally, we give the conclusion in Sec. IV.

%%%%%%%%%%%%%%%%%%%%%%%%%%%%%%%%%%
\section{The SM expectations}
%%%%%%%%%%%%%%%%%%%%%%%%%%%%%%%%%%

\begin{figure}[htbp]
	\centering
	\subfigure[~Within the Standard Model]{
		\label{Feyn-10}
		\includegraphics[width=0.45\textwidth]{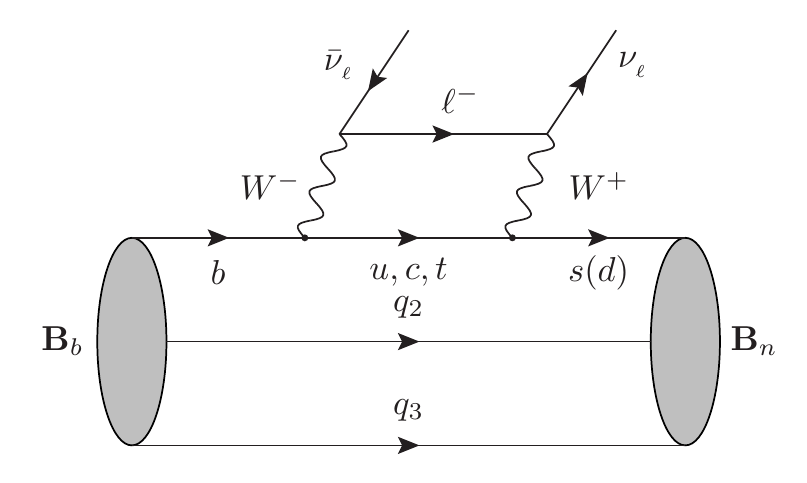}} 
	\hspace{0.5cm}
	\subfigure[~Beyond the Standard Model]{
		\label{Feyn-9}
		\includegraphics[width=0.45\textwidth]{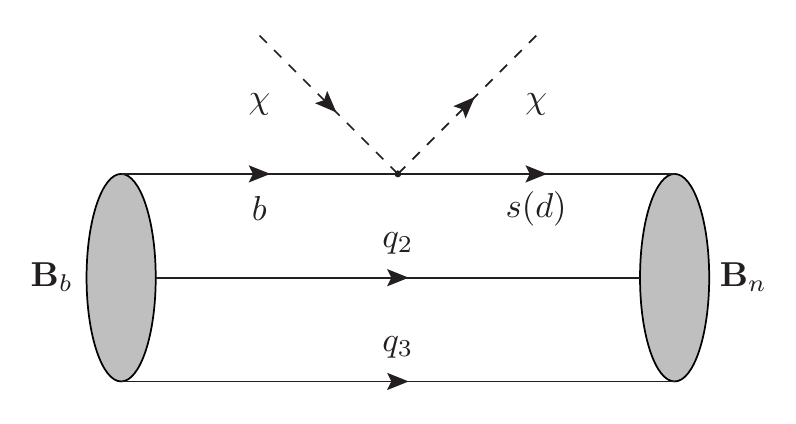}}
	\caption{Feynman diagrams of bottomed baryon FCNC decays with missing energy.}
	\label{Feyn}
\end{figure}

The FCNC decay processes of bottomed baryons with missing energy are described in Fig.~\ref{Feyn}, where ${\bf B}_b$ and ${\bf B}_n$ represent the initial and final baryons, respectively; $q=b$ and $q_{f}=s(d)$ are initial and final quarks, respectively; and $q_{2(3)}$ are the spectator quarks. 

In the SM, there is no tree-level contribution to the FCNC decays of ${\bf B}_b \to {\bf B}_n\bar\nu\nu$. The first-order contributions to these processes come from the penguin and box diagrams as shown in Fig.~\ref{Feyn}a,  which can be described by the effective Lagrangian, given by~\cite{Inami:1980fz}
\begin{equation}
	\begin{aligned}
		\mathcal L_{\bar\nu\nu} = \frac{4G_F}{\sqrt{2}}\frac{\alpha}{2\pi\sin^2\theta_W}\sum_{\ell=e, \mu, \tau}\sum_{q=u, c, t}V_{bq}V_{sq}X^\ell(x_q)(\bar s_{_L}\gamma^\mu b_{_{L}})(\bar\nu_{_{\ell L}}\gamma_\mu\nu_{_{\ell L}}),
		\label{eq25}
	\end{aligned}
\end{equation}
with
\begin{equation}
	\begin{aligned}
		X^\ell(x_q) = \frac{x_q}{8}\left[\frac{x_q+2}{x_q-1}+\frac{3(x_q-2)}{(x_q-1)^2}\ln x_q\right],
		\label{eq26}
	\end{aligned}
\end{equation}
where $G_F$ represents the Fermi coupling constant, $\alpha$ corresponds to the fine structure constant, $\theta_W$ stands for the Weinberg angle, $V_{ij}$ are the Cabibbo–Kobayashi–Maskawa (CKM) matrix elements, and $x_q=m_q^2/M_W^2$ with $m_q$ ($M_W$) being the mass of the quark ($W$-boson). Consequently, the transition amplitude is given by
\begin{equation}
	\begin{aligned}
		\langle {\bf B}_n\bar\nu\nu|\mathcal L_{\bar\nu\nu}| {\bf B}_b\rangle &=\frac{\sqrt{2}G_F\alpha}{4\pi\sin^2\theta_W}V_{bt}V_{st}X^\ell(x_t)\langle {\bf B}_n|\bar s\gamma^\mu(1-\gamma^5)b|{\bf B}_b\rangle\times \bar u_{\nu_\ell}\gamma_\mu(1-\gamma^5)v_{\nu_\ell}.
		\label{eq27}
	\end{aligned}
\end{equation}

The baryonic transition matrix elements can be parameterized by the form factors (FFs) of $f_i^{V,A}~(i=1,2,3)$, $f^S$ and $f^P$, defined by
\begin{equation}
	\begin{aligned}
		\langle {\bf B}_n(P_f,s_f)|(\bar q_{f}\gamma_\mu q) |{\bf B}_b(P,s)\rangle 
		&=\bar u_{ _{{\bf B}_n}} (P_f,s_f)\left[\gamma_\mu f^V_1(q^2) + i\sigma_{\mu\nu} \frac{q^\nu}{M} f^V_2(q^2)+\frac{q^\mu}{M}f^V_3(q^2)\right]u_{_{{\bf B}_b}}(P,s),\\
		\langle {\bf B}_n(P_f,s_f)|(\bar q_{f} q) |{\bf B}_b(P,s)\rangle 
		&=\bar u_{_{{\bf B}_n}} (P_f,s_f)f^S(q^2)u_{_{{\bf B}_b}}(P,s),\\
		\langle {\bf B}_n(P_f,s_f)|(\bar q_{f}\gamma_\mu \gamma^5 q) |{\bf B}_b(P,s)\rangle
		&=\bar u_{_{{\bf B}_n}} (P_f,s_f)\left[\gamma_\mu f^A_1(q^2) + i\sigma_{\mu\nu} \frac{q^\nu}{M} f^A_2(q^2)+\frac{q^\mu}{M}f^A_3(q^2)\right]\gamma^5u_{_{{\bf B}_b}}(P,s),\\
		\langle {\bf B}_n(P_f,s_f)|(\bar q_{f} \gamma^5 q) |{\bf B}_b(P,s)\rangle 
		&=\bar u_{_{{\bf B}_n}} (P_f,s_f)f^P(q^2)\gamma^5 u_{_{{\bf B}_b}}(P,s), 
		\label{ffb}
	\end{aligned}
\end{equation}
where $q$ corresponds to the momentum transfer, and $M$ is the mass of the initial baryon. We will evaluate these elements in terms of the MBM,  which works well for the heavy baryonic decays~\cite{Geng:2020ofy,Geng:2021sxe,Liu:2021rvt,Liu:2022bdq,Liu:2022pdk}. In the MBM, the baryon wave functions at rest are read as 
\begin{equation}\label{wave_function_inrest}
	\Psi(x_{q_1},x_{q_2},x_{q_3}) ={\cal N}\int 
	d^3 \vec{x}
	\prod_{i=1,2,3}
	\phi_{q_i}(\vec{x}_{q_i} - \vec{x}) e^{-iE_{q_i}t_{q_i}} \,,
\end{equation}
where $q_i$ are the quark components of the baryons, ${\cal N}$ the overall normalization constant, $x_{q_i}$ ($E_{q_i}$) the spacetime coordinates (energies) of $q_i$, and $\phi_{q_i}(x)$  the quark wave functions inside a static bag, located at the center, given by 
\begin{equation}\label{quark_wave_function}
	\phi_{q}(\vec{x})  = \left(
	\begin{array}{c}
		\omega_{q+} j_0(p_qr) \chi_q\\
		i\omega_{q-} j_1(p_qr) \hat{r} \cdot \vec{\sigma} \chi_q\\
	\end{array}
	\right)\,.
\end{equation}
Here,  $j_{0,1}$ represent the spherical Bessel functions,  $\omega_{q\pm} = \sqrt{T_{q} \pm M_{q} }$   with $T_{q}$ the kinematic energies,  and $\chi_q$ are  the two component spinors. 
By demanding that quark currents shall not penetrate the boundary of   bags,  we have the boundary condition
\begin{equation}\label{momentum_condition}
	\tan (p_qR) = \frac{p_qR}{1-M_qR-E_qR}\,,
\end{equation}
where $R$ is the bag radius, resulting in that  the magnitudes of  3-momenta are quantized, which can be analogous to the well-know infinite square well.

Several remarks are in order to address some of the issues in the bag model.
One of the main theoretical inconsistencies  is that the chiral symmetry is broken by the  boundary even when the quarks are massless. It is due to that only the  three-momenta are flipped when the quarks meet the boundary, whereas the spin directions are unchanged. Thus, the  boundary inevitably alters the handedness of the quarks. The chiral symmetry plays an important role in the light quark system. Nonetheless, as we only consider the $b\to s$ transitions, of which the chiral symmetry is already broken badly by the $b$ quark mass, it shall not cause severe problems. On the other hand, 
the bag model originally describes a baryon state at rest. Therefore, the form factors at the maxima recoil point~$(q_{max}^2 = (M_{\Lambda_b} - M_\Lambda )^2)$ would be  more reliable. 
In particular, the axial form factors of the $n\to p$ transition is found to be $f_1^A = 1.31$, which is very close to $1.27$ from the experiments.

\begin{figure}[h]
	\centering
	\subfigure[]{
		\label{ff-11}
		\includegraphics[width=0.45\textwidth]{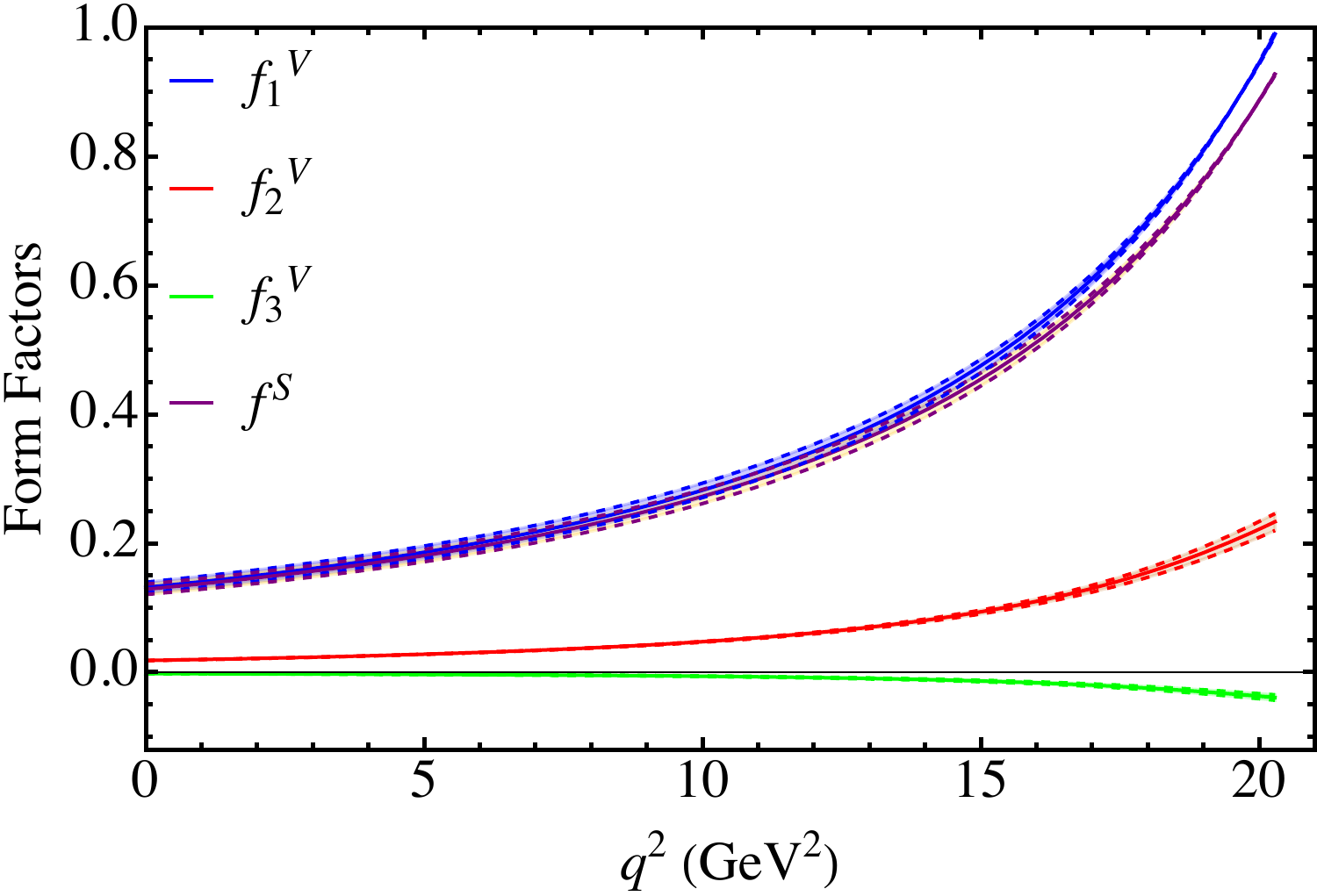}} 
	\hspace{0.5cm}
	\subfigure[]{
		\label{ff-12}
		\includegraphics[width=0.45\textwidth]{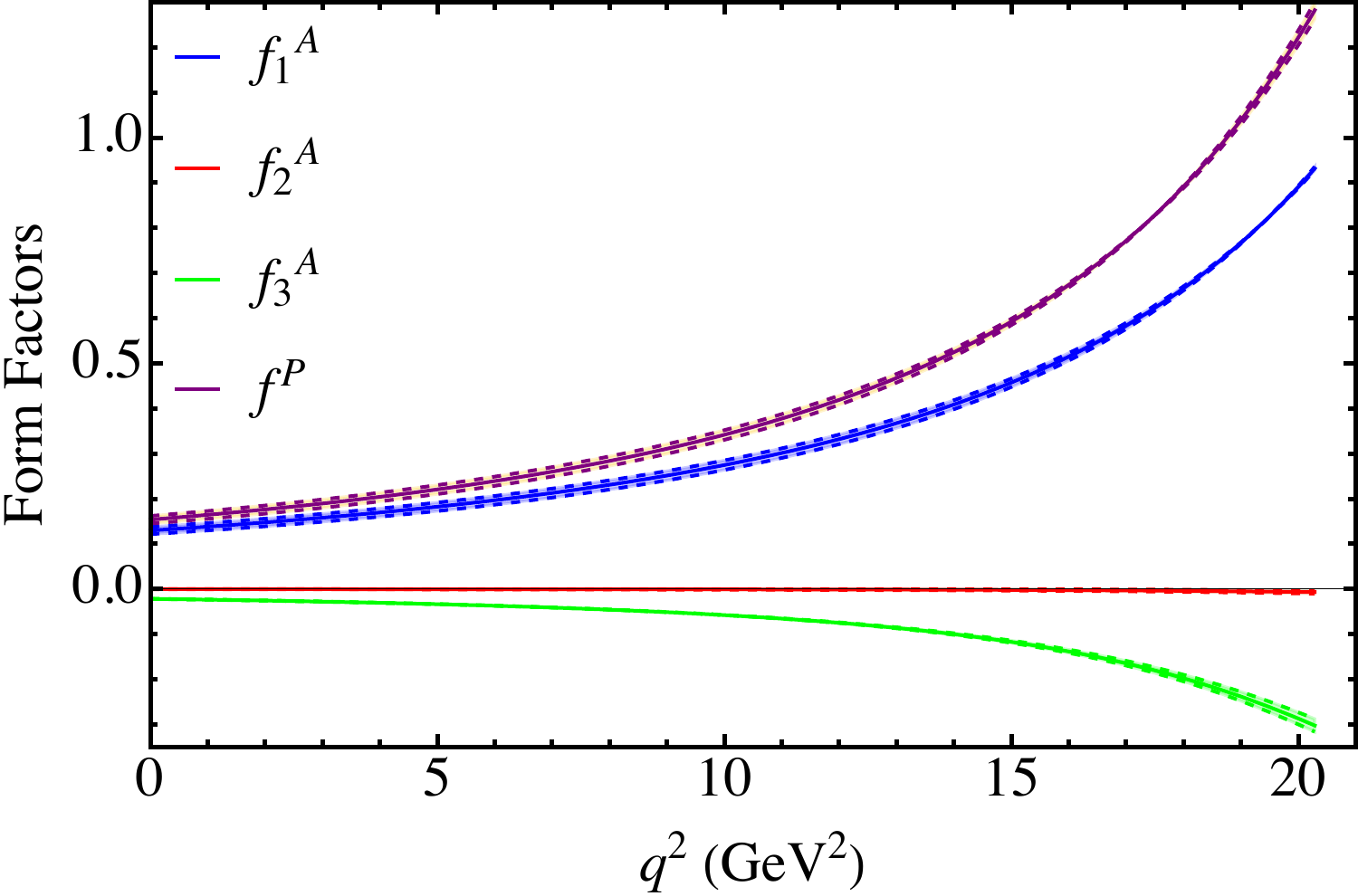}}  \\
	\caption{Form factors of $\Lambda_b\to \Lambda$ as functions of $q^2$}
	\label{ff-1}
\end{figure}

\begin{figure}[h]
	\centering
	\subfigure[]{
		\label{ff-21}
		\includegraphics[width=0.45\textwidth]{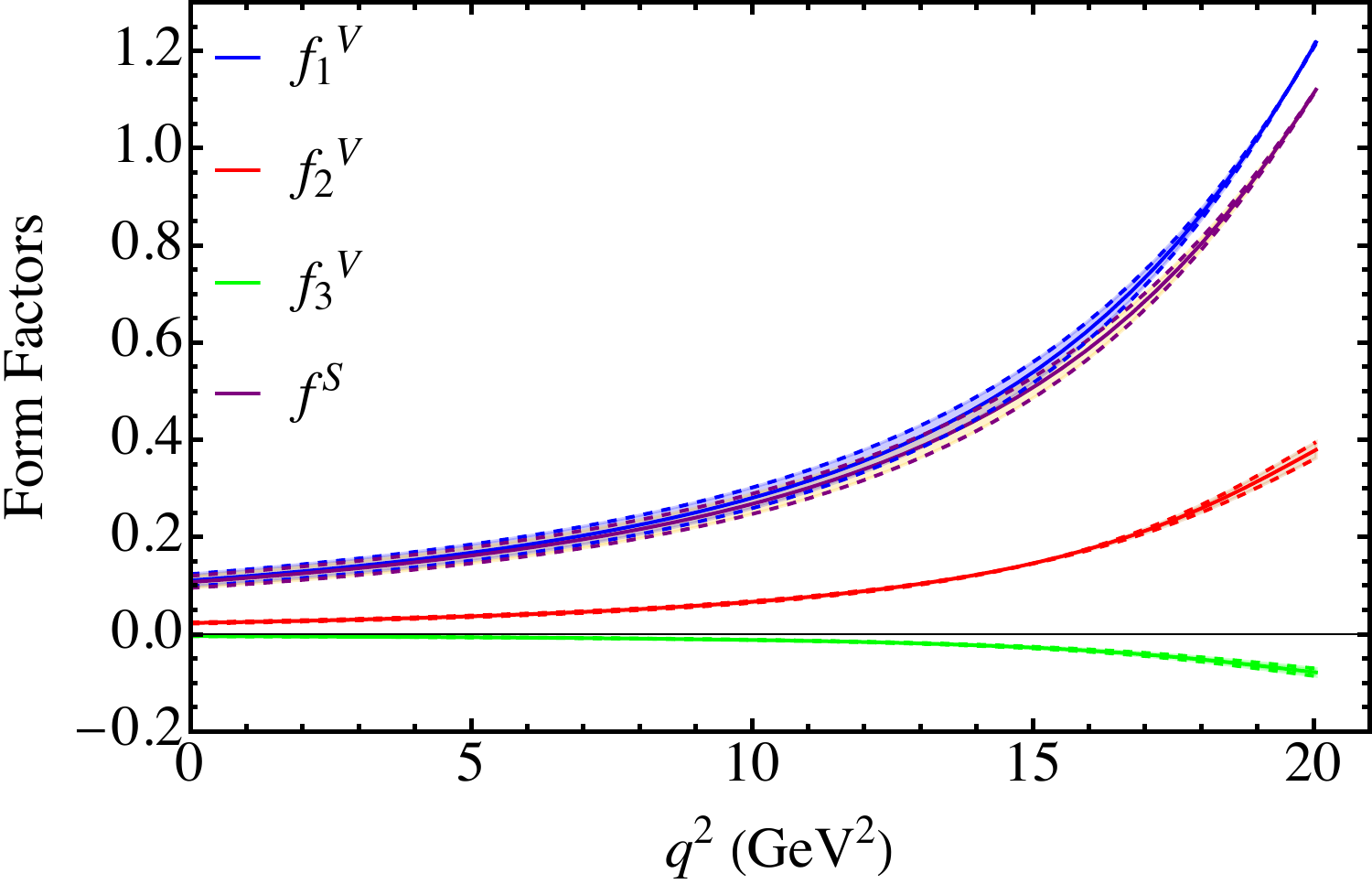}} 
	\hspace{0.5cm}
	\subfigure[]{
		\label{ff-22}
		\includegraphics[width=0.45\textwidth]{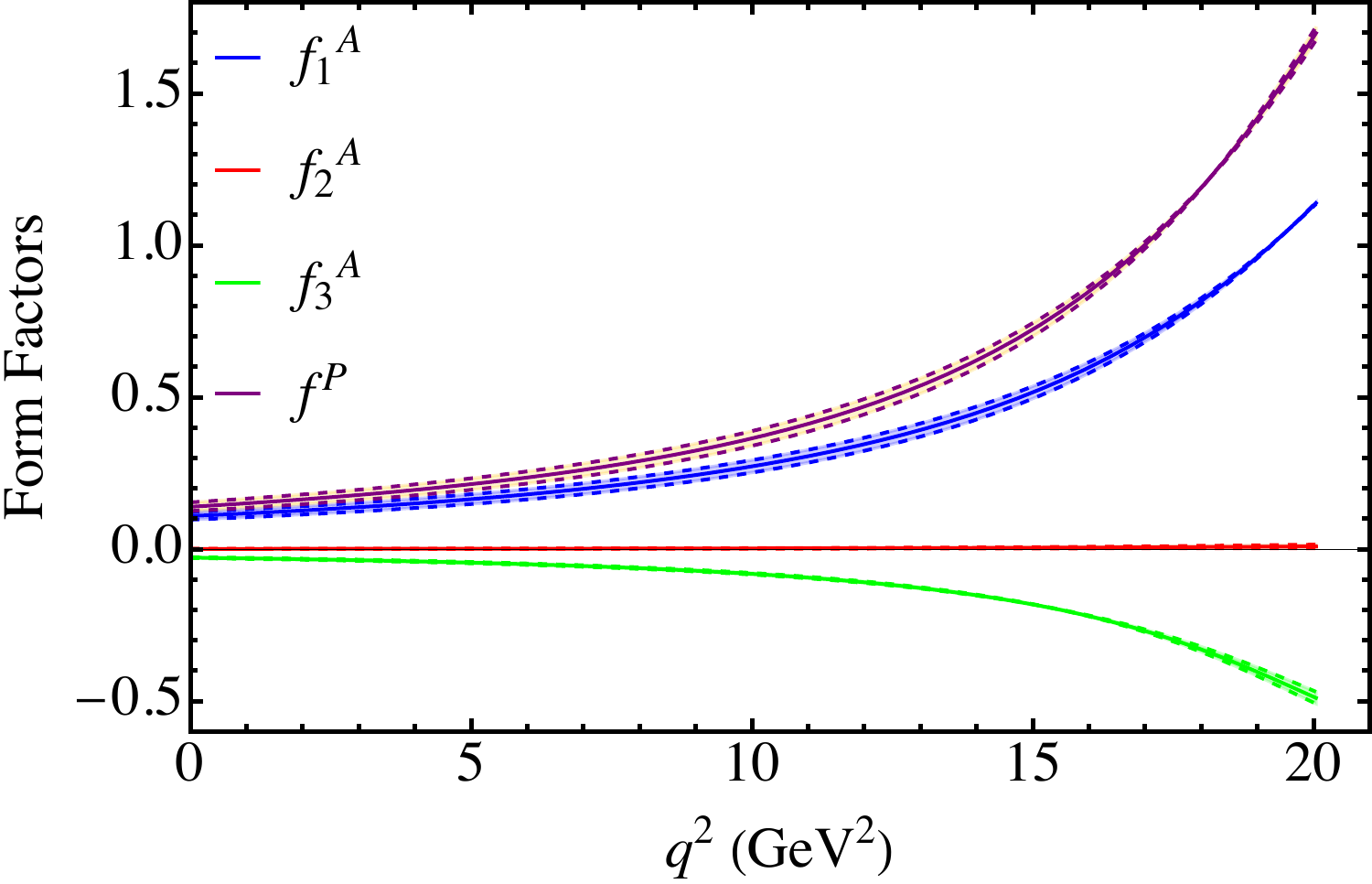}}  \\
	\caption{Form factors of $\Xi_b^{0(-)}\to \Xi^{0(-)}$ as functions of $q^2$}
	\label{ff-2}
\end{figure}

\begin{figure}[h]
	\centering
	\subfigure[]{
		\label{ff-31}
		\includegraphics[width=0.45\textwidth]{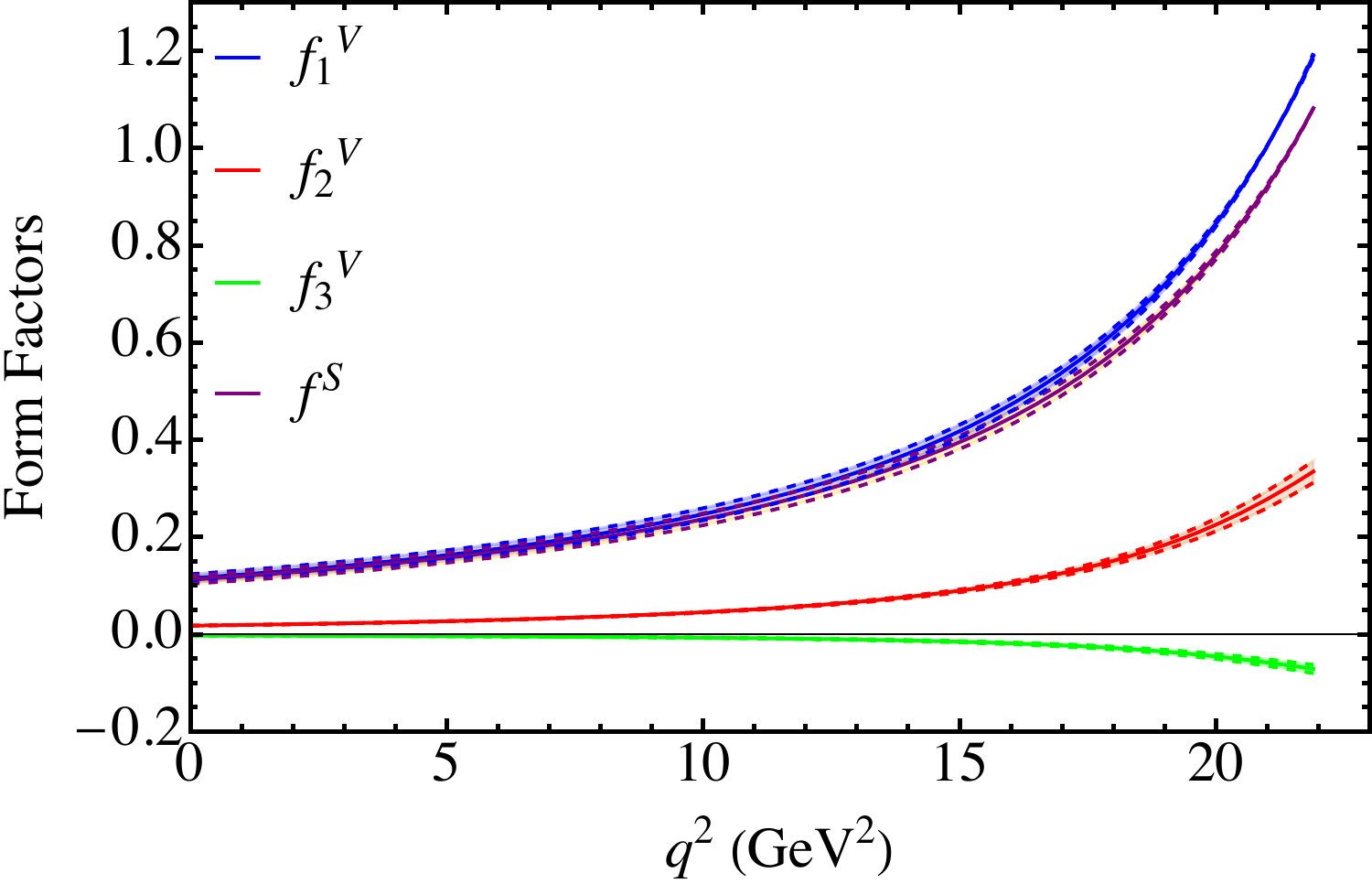}} 
	\hspace{0.5cm}
	\subfigure[]{
		\label{ff-32}
		\includegraphics[width=0.45\textwidth]{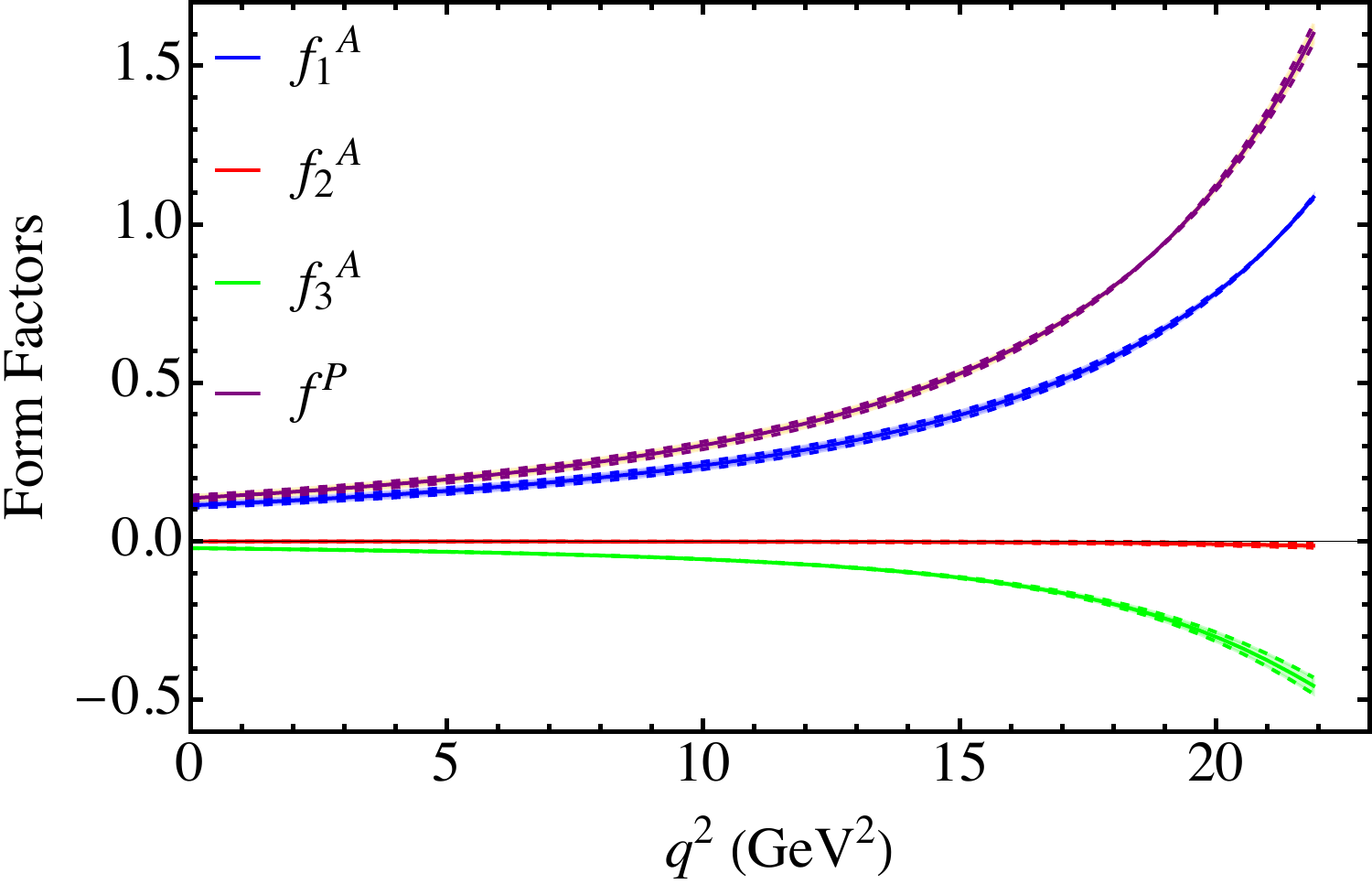}}  \\
	\caption{Form factors of $\Lambda_b\to n$ as functions of $q^2$}
	\label{ff-3}
\end{figure}

\begin{figure}[h]
	\centering
	\subfigure[]{
		\label{ff-41}
		\includegraphics[width=0.45\textwidth]{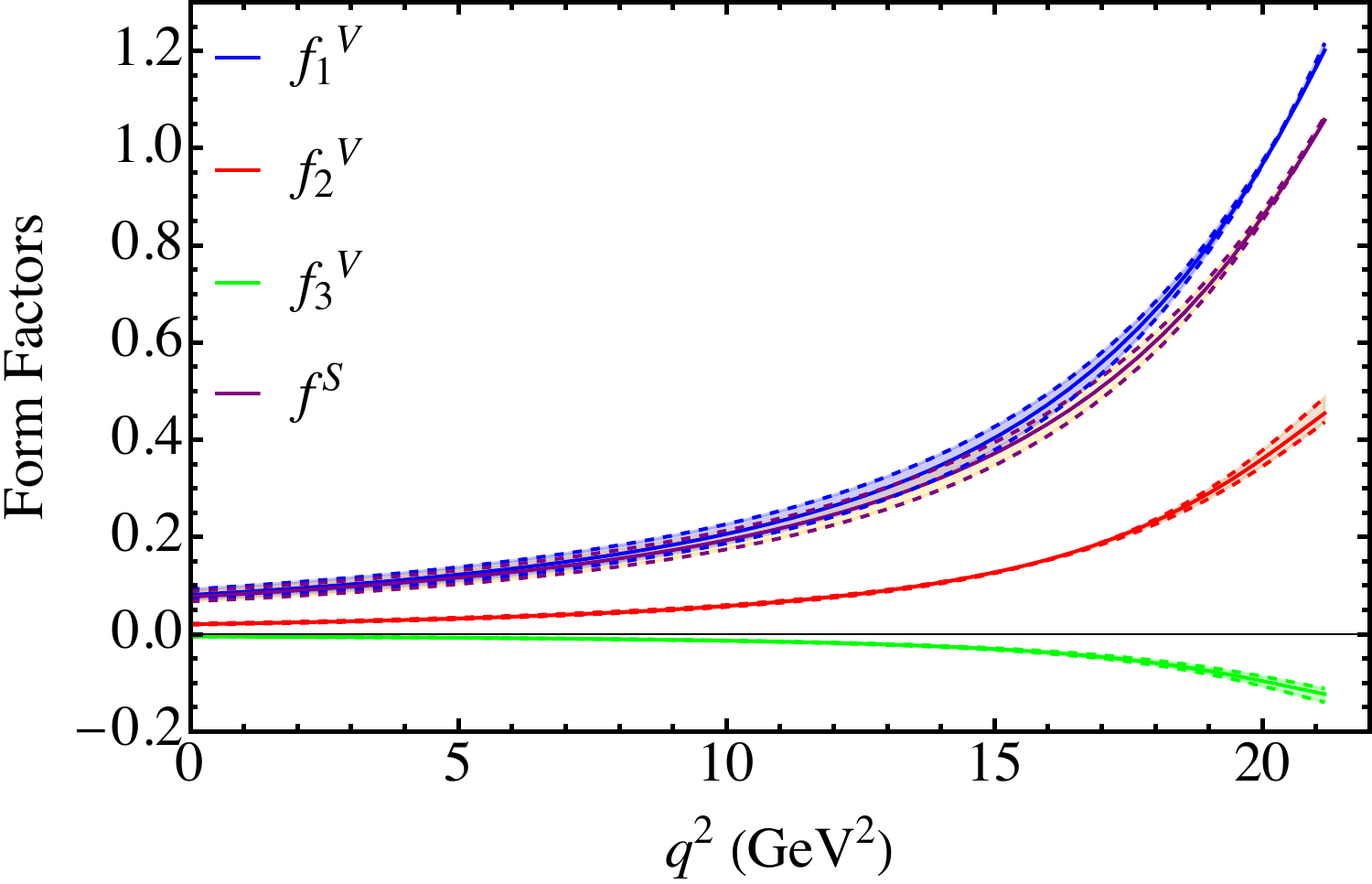}} 
	\hspace{0.5cm}
	\subfigure[]{
		\label{ff-42}
		\includegraphics[width=0.45\textwidth]{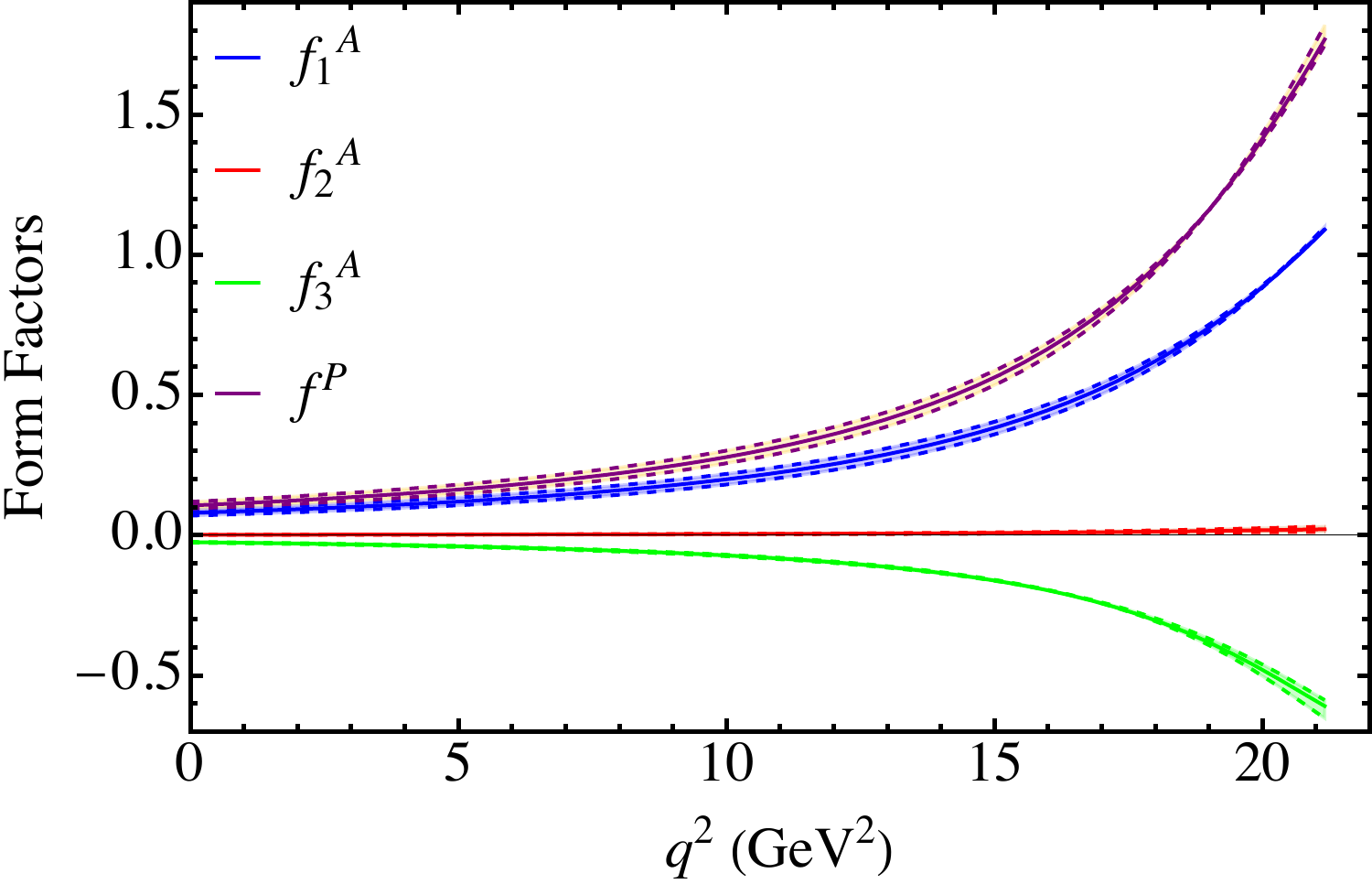}}  \\
	\caption{Form factors of $\Xi_b^-\to \Sigma^-$ as functions of $q^2$}
	\label{ff-4}
\end{figure}

\begin{figure}[h]
	\centering
	\subfigure[]{
		\label{ff-51}
		\includegraphics[width=0.45\textwidth]{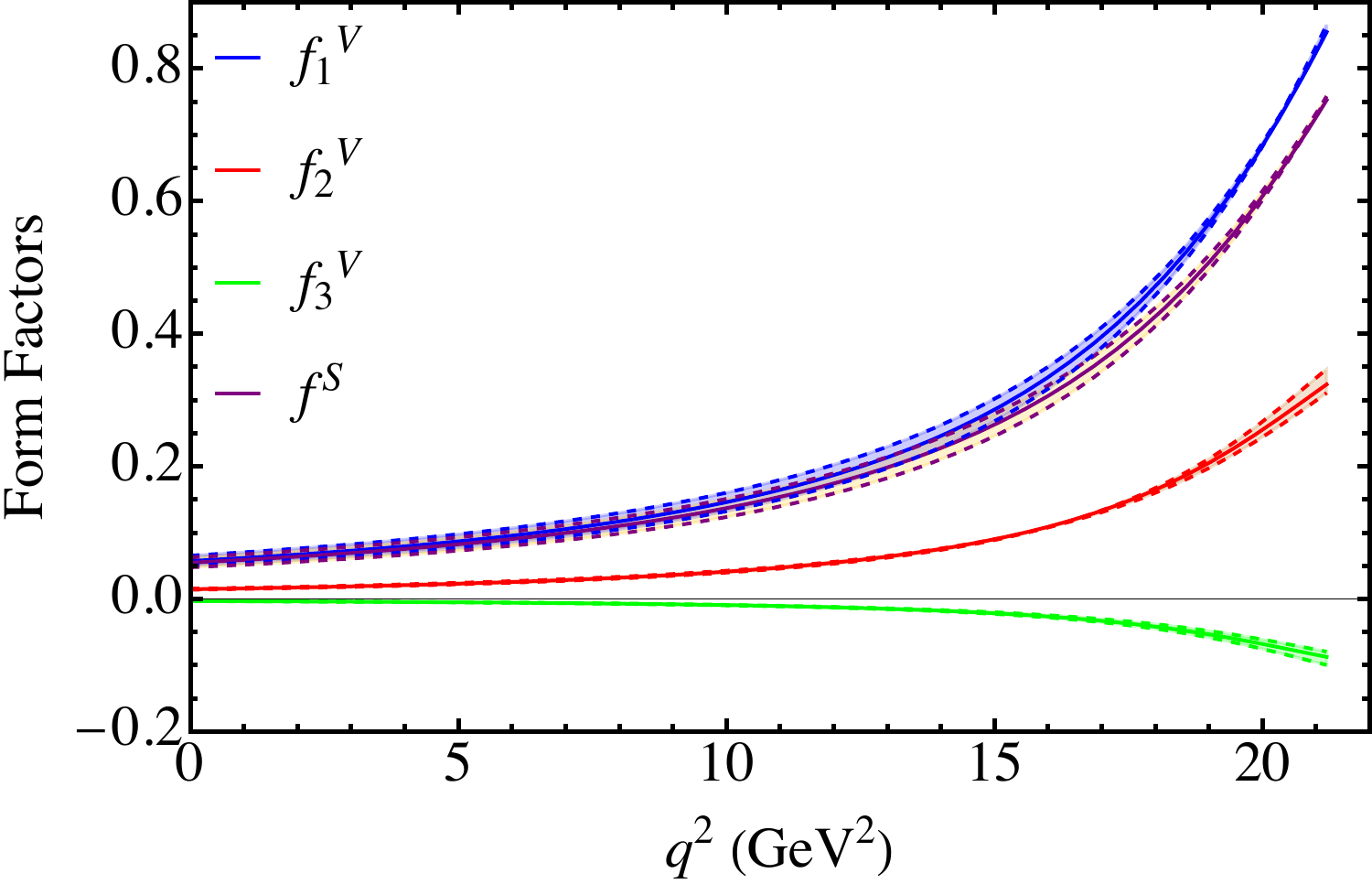}} 
	\hspace{0.5cm}
	\subfigure[]{
		\label{ff-52}
		\includegraphics[width=0.45\textwidth]{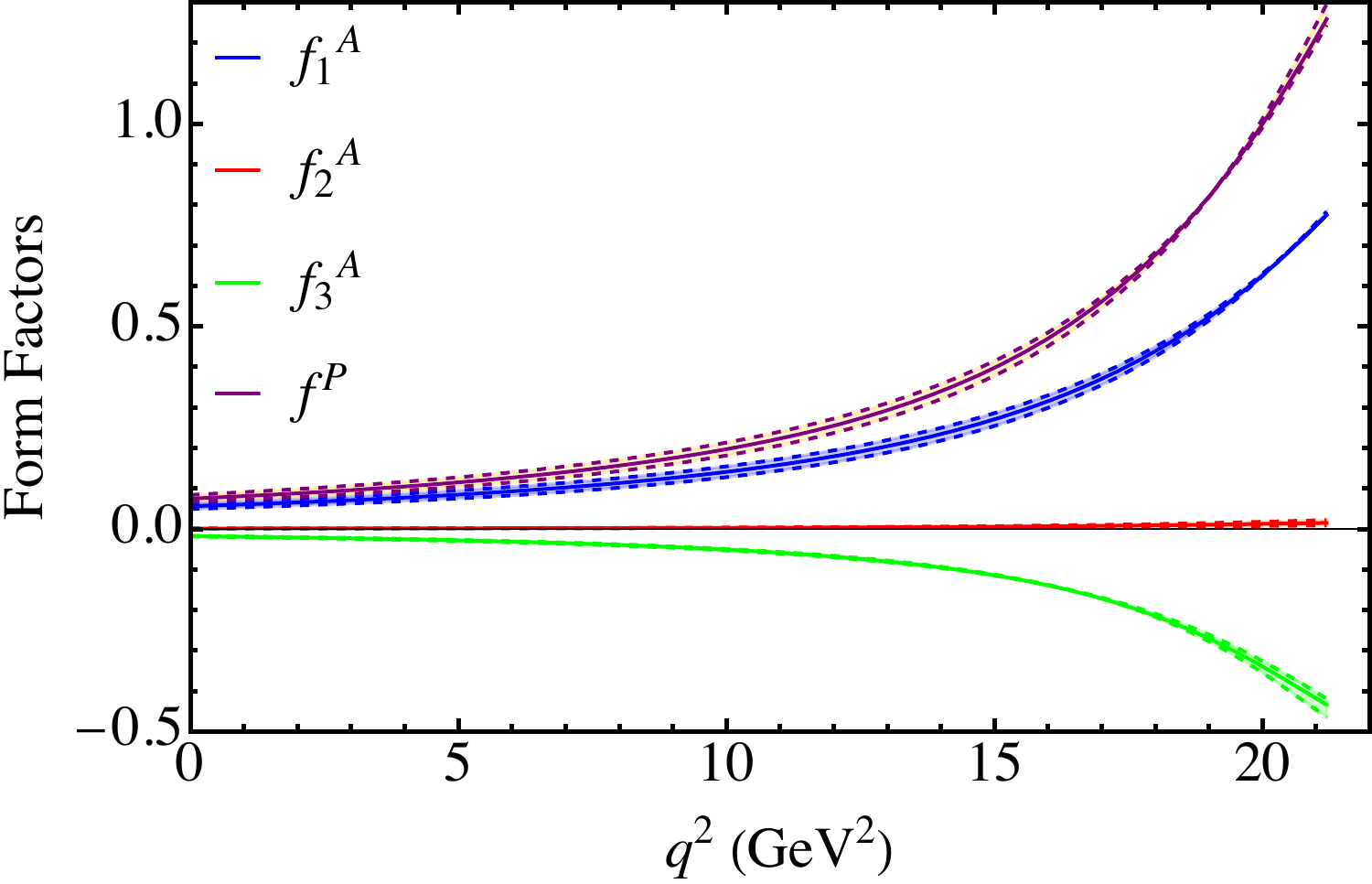}}  \\
	\caption{Form factors of $\Xi_b^0\to \Sigma^0$ as functions of $q^2$}
	\label{ff-5}
\end{figure}

\begin{figure}[h]
	\centering
	\subfigure[]{
		\label{ff-61}
		\includegraphics[width=0.45\textwidth]{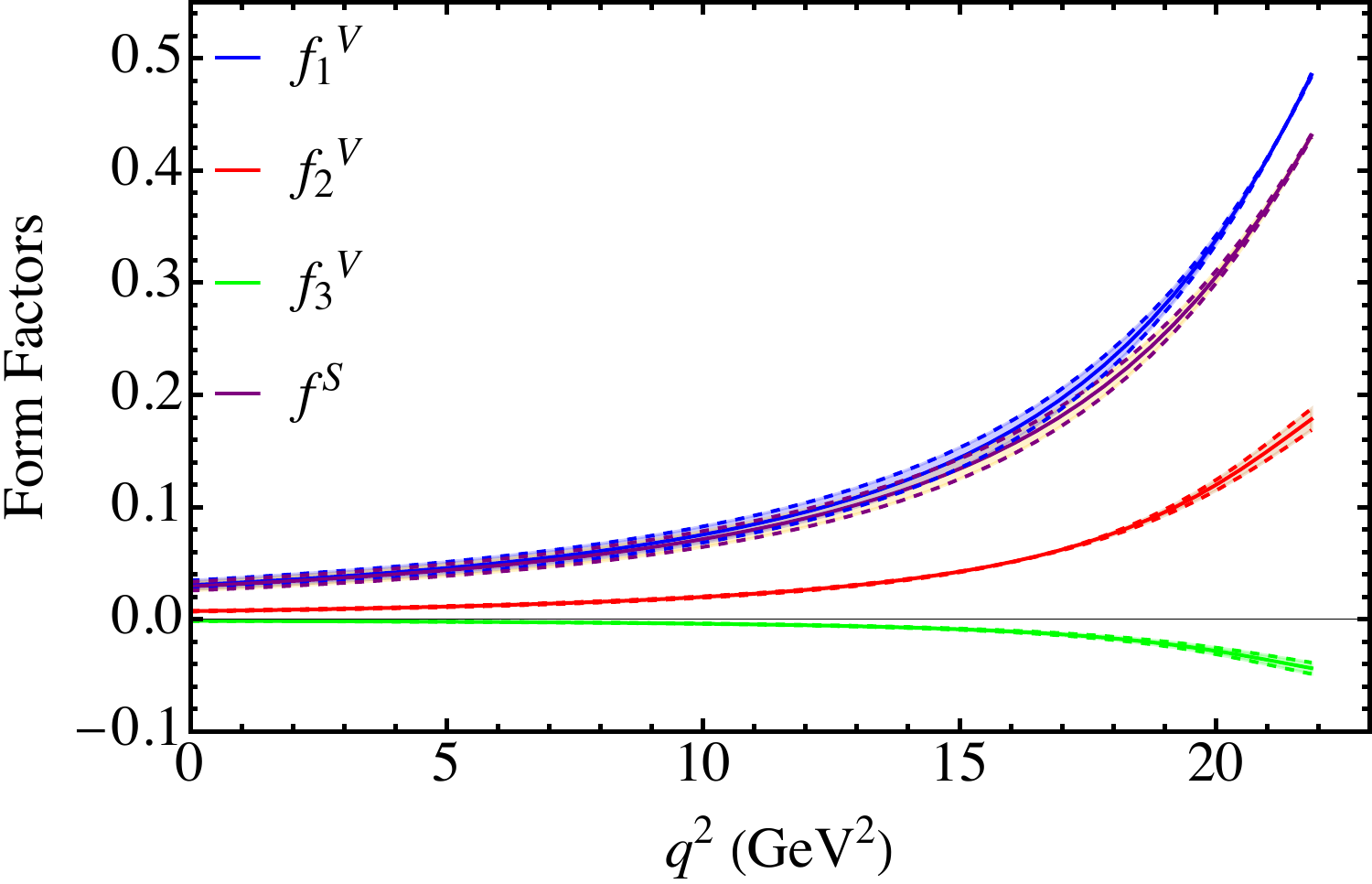}} 
	\hspace{0.5cm}
	\subfigure[]{
		\label{ff-62}
		\includegraphics[width=0.45\textwidth]{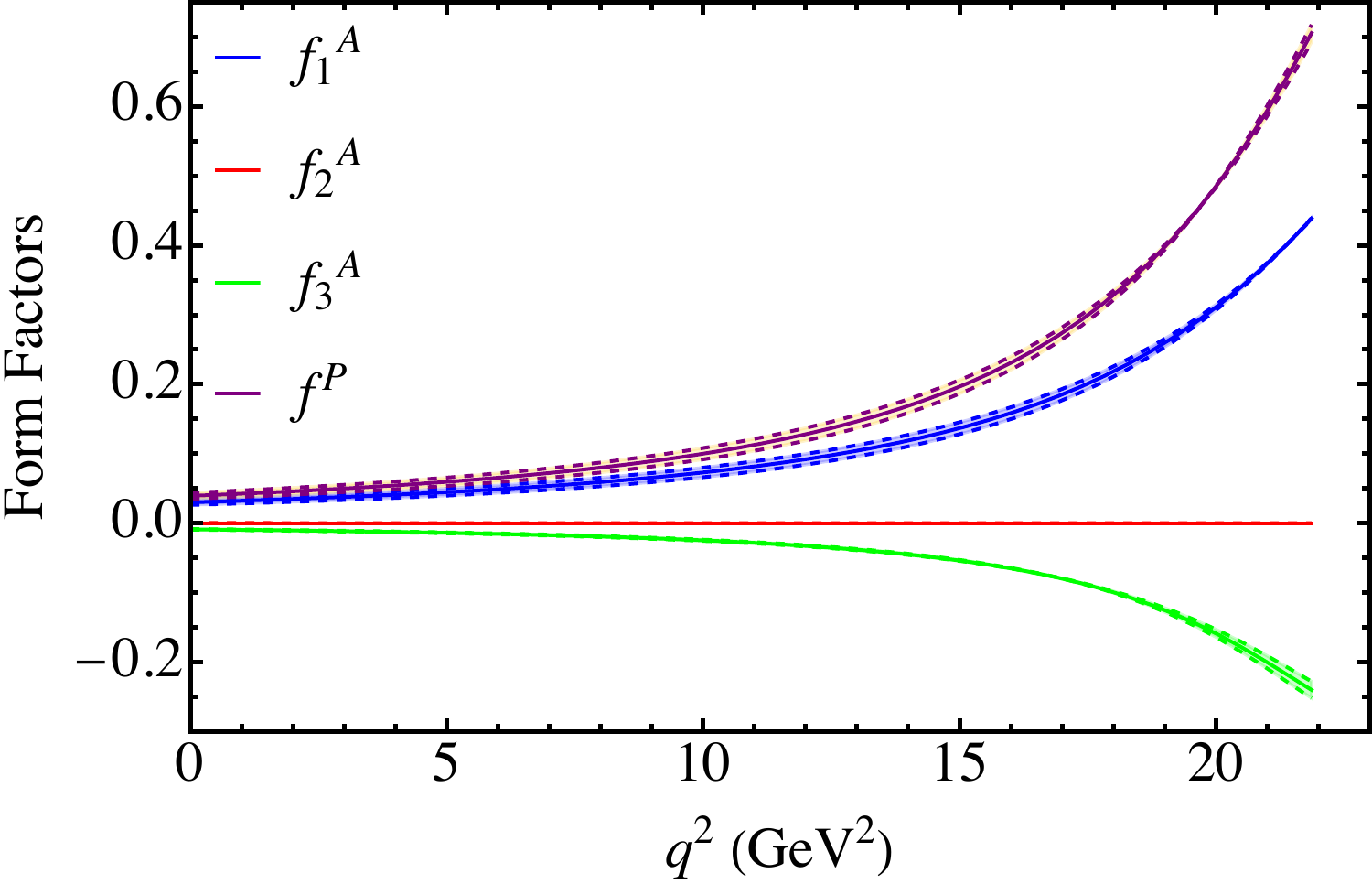}}  \\
	\caption{Form factors of $\Xi_b^0\to \Lambda$ as functions of $q^2$}
	\label{ff-6}
\end{figure}

By sandwiching the operators, we arrive 
\begin{eqnarray}\label{master}
	%&&\int \langle B_f | \overline{ q } b(x) e^{iq_Mx} |\Lambda_b \rangle d^4 x=\nonumber\\
	&&\int \langle \Lambda | \overline{ s }\Gamma  b(x) e^{iqx} |\Lambda_b \rangle d^4 x= {\cal Z}\int d^3\vec{x}_\Delta \Gamma _{sb}(\vec{x}_\Delta) \prod_{q_j=u,d } D_{q_j}(\vec{x}_\Delta)\,,
\end{eqnarray}
with
\begin{eqnarray}\label{con}
	&&{\cal Z} \equiv (2\pi )^4 \delta^4(p_{\Lambda_b} - p_{\Lambda} -q)    {\cal N}_{\Lambda_b} {\cal N}_{\Lambda} \,,\nonumber\\
	&&D_{q_j}(\vec{x}_{\Delta}) \equiv \sqrt{1- v^2 } \int d^3 \vec{x} \phi_{q_j}^\dagger \left(\vec{x} +\frac{1}{2}\vec{x}_{\Delta}\right) \phi_{q_j} \left(\vec{x} -\frac{1}{2}\vec{x}_{\Delta}\right)
	e^{-2iE_{q_j}  \vec{v}\cdot\vec{x}}\,,\nonumber\\
	&&\Gamma _{sb} (\vec{ x}_\Delta)=\int  d^3\vec{x}  \phi _s\left(\vec{x} + \frac{1}{2}\vec{x}_\Delta \right)\gamma^0 S_{-\vec{ v}}\Gamma S_{-\vec{ v}} \phi_b\left(\vec{x} - \frac{1}{2}\vec{x}_\Delta \right) e^{i(M_\Lambda  + M_{\Lambda_b}-E_s-E_b)\vec{ v}\cdot\vec{ x}     }\,,
\end{eqnarray}
where $\Gamma$ are arbitrary Dirac matrices, and $S_{\vec{v}}$ the Lorentz boost matrix of Dirac spinors.  We have taken the initial (final) state as $\Lambda_b$ ($\Lambda$) for an concrete example. To simplify the algebra,  
the Briet frame is chosen, where $\Lambda_b$ and $\Lambda$ have the velocity $-\vec{ v}$ and $\vec{ v}$, respectively. Notably, all the parameters of  the model are extracted from the mass spectra, given as~\cite{Zhang:2021yul}
\begin{equation}
	R = 4.8 ~\text{GeV}^{-1}\,,~~~M_{u,d} =0\,,~~~M_s = 0.28~\text{GeV}\,,~~~M_b = 5.093~\text{GeV}\,.
\end{equation}
In general, the bag radius of $\Lambda_b$ differs from the one of  $\Lambda$. Nevertheless, in calculating  the transition matrix elements,  the different bag radii between the initial and final states lead to several theoretical difficulties. In this work, we take  the bag radii of the  initial and final baryons as the same and allow it to vary $5\%$, which shall cover the reasonable range. We consider the bottomed baryon decays of ($\Lambda_b\to\Lambda\bar\nu\nu$ and $\Xi_b^{0(-)}\to\Xi^{0(-)}\bar\nu\nu$) and ($\Lambda_b\to n \bar\nu\nu$, $\Xi_b^{0(-)}\to \Sigma^{0(-)} \bar\nu\nu$, and $\Xi_b^{0}\to \Lambda \bar\nu\nu$), due to the ($b\to s$) and ($b\to d$) transitions at quark level, respectively. The FFs can be extracted straightforwardly after the computations, which are shown in Figs.~{\ref{ff-1}-\ref{ff-6}} along with their $q^2$ dependencies, where the solid lines represent the central values, and the shadows between the dashed lines correspond to the errors estimated by varying the bag radius of $R=4.8 ~\text{GeV}^{-1}$ within $\pm5\%$. 

By integrating the three-body phase space, we obtain the decay branching ratio to be
\begin{equation}
	\mathcal {B} ({\bf B}_b\to{\bf B}_n\bar\nu\nu) = \frac{1}{512  \pi^3 M^3 \Gamma_{{\bf B}_b}}\int\frac {dq^2}{q^2} \lambda^{1/2}(M^2, q^2, M_f^2)\lambda^{1/2}(q^2, m_1^2, m_2^2)\int d\cos\theta\sum|\mathcal M|^2,
	\label{ps3}
\end{equation}
where $\lambda(x, y, z)= x^2 + y^2 +z^2 -2xy-2xz -2yz$ is the K${\rm \ddot a}$llen function, $M$, $M_f$, $m_1$ and $m_2$ correspond to the masses of the initial baryon, final baryon, neutrino and anti-neutrino, respectively, $\theta$ is the phase space angle, $\Gamma_{{\bf B}_b}$ represents the total width of the initial baryon, and $\mathcal M$ stands for the amplitude. As the three generations of neutrinos are indistinguishable experimentally, the final results need to be multiplied by three. For the $b\to s$ transition, the decay branching ratios associated with neutrino and anti-neutrino pairs are as follows:
\begin{equation}
	\begin{aligned}
		\mathcal{B}(\Lambda_b\to \Lambda\bar\nu\nu)&=5.52_{-0.28}^{+0.28}\times 10^{-6}, \\
		\mathcal{B}(\Xi_b^{0(-)}\to \Xi^{0(-)}\bar\nu\nu)&=7.80^{+0.71}_{-0.67}\times 10^{-6}.
		\label{SM1}
	\end{aligned}
\end{equation}
Here, due to the $\rm SU(3)$ flavor symmetry, the branching ratios of $\Xi_b^0$ and $\Xi_b^-$ are considered approximately to be equal. The uncertainties of $\mathcal B$ are about $\pm5\%$ to $\pm10\%$. Note that our results of ($\Lambda_b\to \Lambda\bar\nu\nu$) in Eq.~\eqref{SM1} is smaller than the previous prediction in Ref.~\cite{Chen:2000mr}. Similarly, for the $b\to d$ transition we have that
\begin{equation}
	\begin{aligned}
		\mathcal{B}(\Lambda_b\to n\bar\nu\nu)&=2.76_{-0.16}^{+0.17}\times 10^{-7}, \\
		\mathcal{B}(\Xi_b^-\to \Sigma^-\bar\nu\nu)&=2.65_{-0.26}^{+0.29}\times 10^{-7}, \\
		\mathcal{B}(\Xi_b^0\to \Sigma^0\bar\nu\nu)&=1.24_{-0.12}^{+0.13}\times 10^{-7},  \\
		\mathcal{B}(\Xi_b^0\to \Lambda\bar\nu\nu)&=3.88_{-0.40}^{+0.37}\times 10^{-8}, 
		\label{SM2}
	\end{aligned}
\end{equation}
which are about one to two orders of magnitude smaller than the modes in Eq.~\eqref{SM1}, due to the ratio of the Cabibbo–Kobayashi–Maskawa (CKM) matrix elements, $|V_{td}/V_{ts}|\sim\mathcal O(\lambda)$. As a result, the SM predictions of bottomed baryonic FCNC processes with $\slashed E$ are $\mathcal O(10^{-8})-\mathcal O (10^{-6})$. If the experimental detections of these decays are larger than the values in Eqs.~\eqref{SM1}-\eqref{SM2}, the new invisible neutral particles from NP are expected. 

%%%%%%%%%%%%%%%%%%%%%%%%%%%%%%%%%%
\section{Processes with invisible particles}
%%%%%%%%%%%%%%%%%%%%%%%%%%%%%%%%%%

\subsection{Effective Lagrangian}
In Fig.~\ref{Feyn}b, two spin-$1/2$ invisible Majorana particles of $\chi\chi$ are assumed to be emitted in the process, in which the four-fermion vertex may be generated at tree or loop level by introducing new physical mediators in specific models~\cite{Matsumoto:2018acr,Yang:2016wrl,Chala:2015ama}. Under the low energy scale, the model-independent effective Lagrangian is given by
\begin{equation}
\begin{aligned}
\mathcal L_{eff}=\sum_{i=1}^{6}{g_{_{mi}} Q_i},
\label{eq1}
\end{aligned}
\end{equation}
where $g_{fi}$ are the phenomenological coupling constants, which are taken at the new physical energy scale $\Lambda$. There are $6$ independent dimension-six effective operators, which have the forms: 
\begin{equation}
\begin{aligned}
& Q_1=(\bar q_{_f}q)(\chi\chi),~~~~~~~~~Q_2=(\bar q_{_f} \gamma^{5}q)(\chi\chi),~~~~~~~~~ Q_3=(\bar q_{_f}q)(\chi\gamma^{5}\chi),\\
& Q_4=(\bar q_{_f} \gamma^{5}q)(\chi\gamma^{5}\chi),~~~Q_5=(\bar q_{_f} \gamma_{\mu}q)(\chi\gamma^{\mu}\gamma^{5}\chi),~~~ Q_6=(\bar q_{_f} \gamma_{\mu}\gamma^{5} q)(\chi\gamma^{\mu}\gamma^{5}\chi),  
\label{op}
\end{aligned}
\end{equation}
where the invisible particles of $\chi$ have been assumed to be the Majorana type. Since $\chi\gamma^{\mu}\chi=0$ and $\chi\sigma^{\mu\nu}\chi=0$, there is no contribution from the vector or tensor current. 

\begin{figure}[htb]
	\centering
	\includegraphics[width=0.45\textwidth]{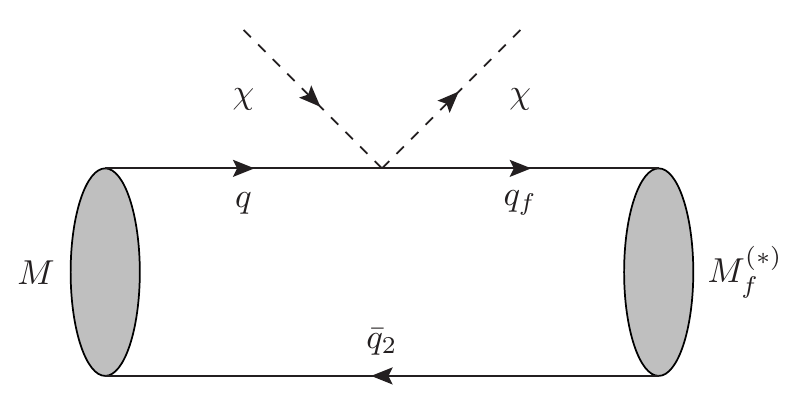}
	\caption{Feynman diagram of bottomed meson FCNC decays with invisible particles.}
	\label{Feyn15}
\end{figure}

The upper limits of the coupling constants in the effective Lagrangian can be extracted from the differences between the theoretical predictions and experimental data of the $B$ meson FCNC decays, such as $ B ^ {-} \to K ^ -(K^{\ast-}) + \slashed E $ and $ B ^ {-} \to \pi ^ -(\rho^-) + \slashed E $, of which Feynman diagram is illustrated as Fig.~\ref{Feyn15}. 
For the $0^-\to0^-$ meson decays of $M^-\to M_f^-\chi\chi$, only operators $Q_{1,3,5}$ give the contributions. And for the $0^-\to1^-$ meson decays of $M^-\to M_f^{*-}\chi\chi$ only operators $Q_{2,4,5,6}$ give the contributions. The amplitudes of the $0^-\to0^-$ decays can be simplified as Eq.~(\ref{Lmm0}), and the amplitudes of the $0^-\to1^-$ decays can be simplified as Eq.~(\ref{Lmm1}),
\begin{equation}
	\begin{aligned}
		\langle M_f^-\chi\chi|\mathcal L_{eff}|M^-\rangle&=2g_{m1}  \langle M_f^-|(\bar q_{_f} q) |M^-\rangle\bar u_{\chi}v_{\chi}+2 g_{m3} \langle M_f^-|(\bar q_{_f} q) |M^-\rangle\bar u_{\chi}\gamma^5 v_{\chi}\\
		&+2g_{m5}\langle M_f^-|(\bar q_{_f}\gamma_\mu q) |M^-\rangle\bar u_{\chi}\gamma^{\mu}\gamma^5 v_{\chi}, 
		\label{Lmm0}
	\end{aligned}
\end{equation}
\begin{equation}
	\begin{aligned}
		\langle M_f^{*-}\chi\chi|\mathcal L_{eff}|M^-\rangle&=2g_{m2} \langle M_f^{*-}|(\bar q_{_f}\gamma^5 q) |M^-\rangle\bar u_{\chi}v_{\chi}+2g_{m4}  \langle M_f^{*-}|(\bar q_{_f}\gamma^5 q) |M^-\rangle\bar u_{\chi}\gamma^5 v_{\chi}\\
		&+2g_{m5} \langle M_f^{*-}|(\bar q_{_f}\gamma_\mu q) |M^-\rangle\bar u_{\chi}\gamma^{\mu}\gamma^5 v_{\chi}+2g_{m6} \langle M_f^{*-}|(\bar q_{_f}\gamma_\mu \gamma^5q) |M^-\rangle\bar u_{\chi}\gamma^{\mu}\gamma^5 v_{\chi}. 
		\label{Lmm1}
	\end{aligned}
\end{equation}

The hadronic transition matrix elements can be expressed as
\begin{equation}
\begin{aligned}
 \langle M_f^-|(\bar q_{_f} q) |M^-\rangle 
&= \frac{M^2-M_{f}^2}{m_q-m_{q_{_f}}}f_0 (q^2),\\
\langle M_f^-|(\bar q_{_f}\gamma_\mu q) |M^-\rangle 
&= (P+P_f)_{\mu}f_+ (q^2)+(P-P_f)_{\mu}\frac{M^2-M_{f}^2}{q^2} \big[f_0 (q^2)-f_+ (q^2)\big],\\
\langle M_f^-|(\bar q_{_f}\sigma_{\mu\nu} q) |M^-\rangle
&=i\big[P_{\mu} (P-P_f)_{\nu}-P_{\nu} (P-P_f)_{\mu}\big] \frac{2}{M+M_f} f_T(q^2),
\label{}
\end{aligned}
\end{equation}
and
\begin{equation}
	\begin{aligned}
		\langle M_f^{*-}|(\bar q_{_f}\gamma^5 q) |M^-\rangle 
		&=-i\big[\epsilon \cdot (P-P_f)\big]\frac{2M_f}{m_q+m_{q_{_f}}}A_0(q^2),\\
		\langle M_f^{*-}|(\bar q_{_f}\gamma_\mu\gamma^5 q) |M^-\rangle
		&= i\bigg\{\epsilon_{\mu} (M+M_f)A_1(q^2)-(P+P_f)_{\mu}\frac{\epsilon \cdot (P-P_f)}{M+M_f}A_2(q^2)\\
		&~~~-(P-P_f)_{\mu}  \big[\epsilon \cdot (P-P_f)\big] \frac{2M_f}{q^2} \big[A_3(q^2)-A_0(q^2)\big]\bigg\},\\
		\langle M_f^{*-}|(\bar q_f \gamma_\mu q )|M^-\rangle
		&=\varepsilon _{\mu \nu \rho \sigma} \epsilon ^\nu P^\rho (P-P_f)^\sigma\frac{2 }{M+M_f}V(q^2), 
		\label{}
	\end{aligned}
\end{equation}
where $m_{q_f}$ are the quark masses, $f_j~(j=0,+,T)$, $A_k~(k=0-3)$ and $V$ are the FFs, which are evaluated from the method of the LCSR~\cite{Ball:2004ye,Aliev:2010ki,Straub:2015ica}, and $\epsilon$ is the polarization vector of the final meson with the convention of $\varepsilon^{0123}=1$. 

\begin{figure}[h]
	\centering
	\subfigure[~($b s\chi\chi$) vertex]{
		\label{gm-1}
		\includegraphics[width=0.45\textwidth]{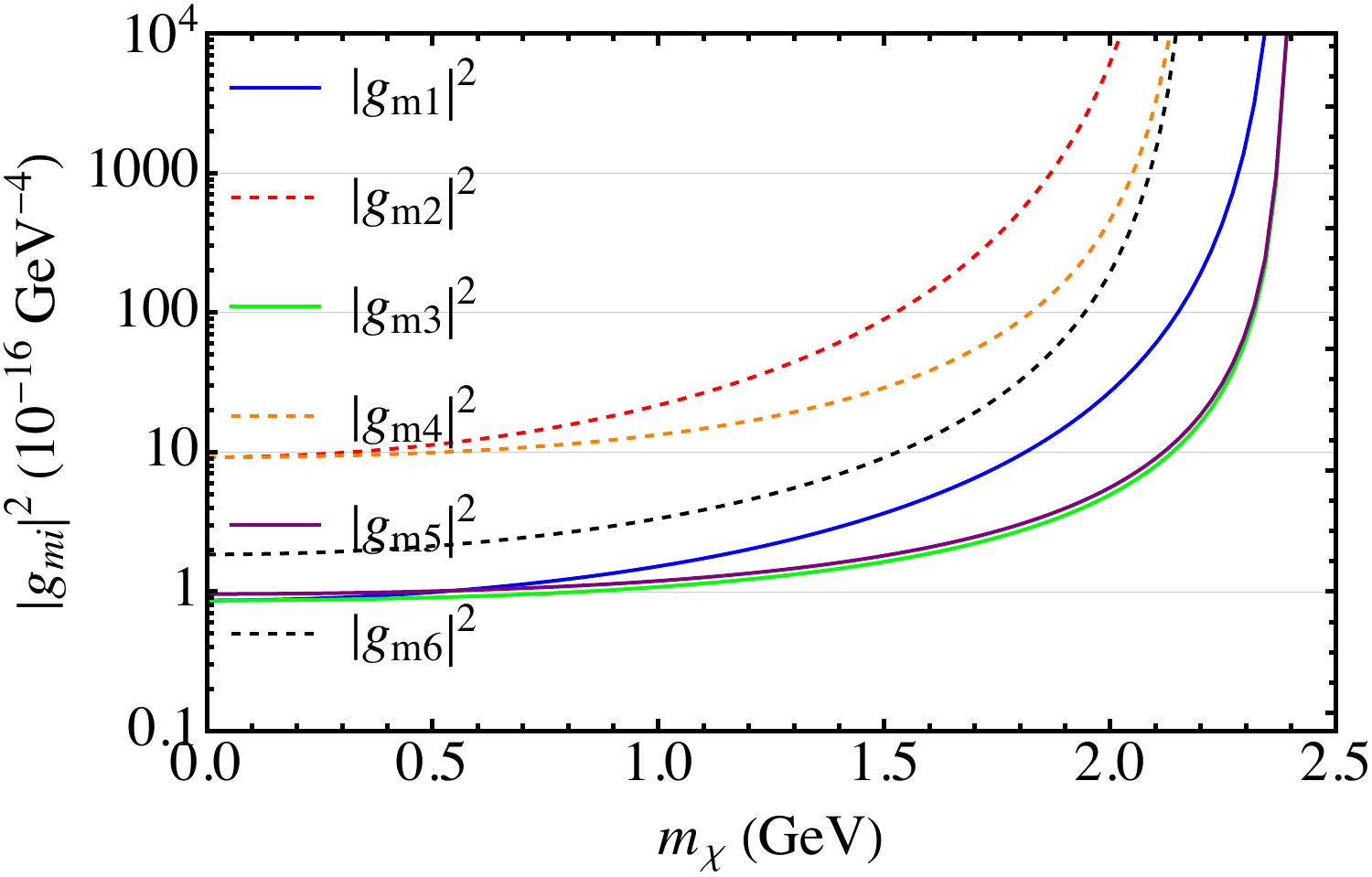}} 
	\hspace{0.5cm}
	\subfigure[~($bd\chi\chi$) vertex]{
		\label{gm-2}
		\includegraphics[width=0.45\textwidth]{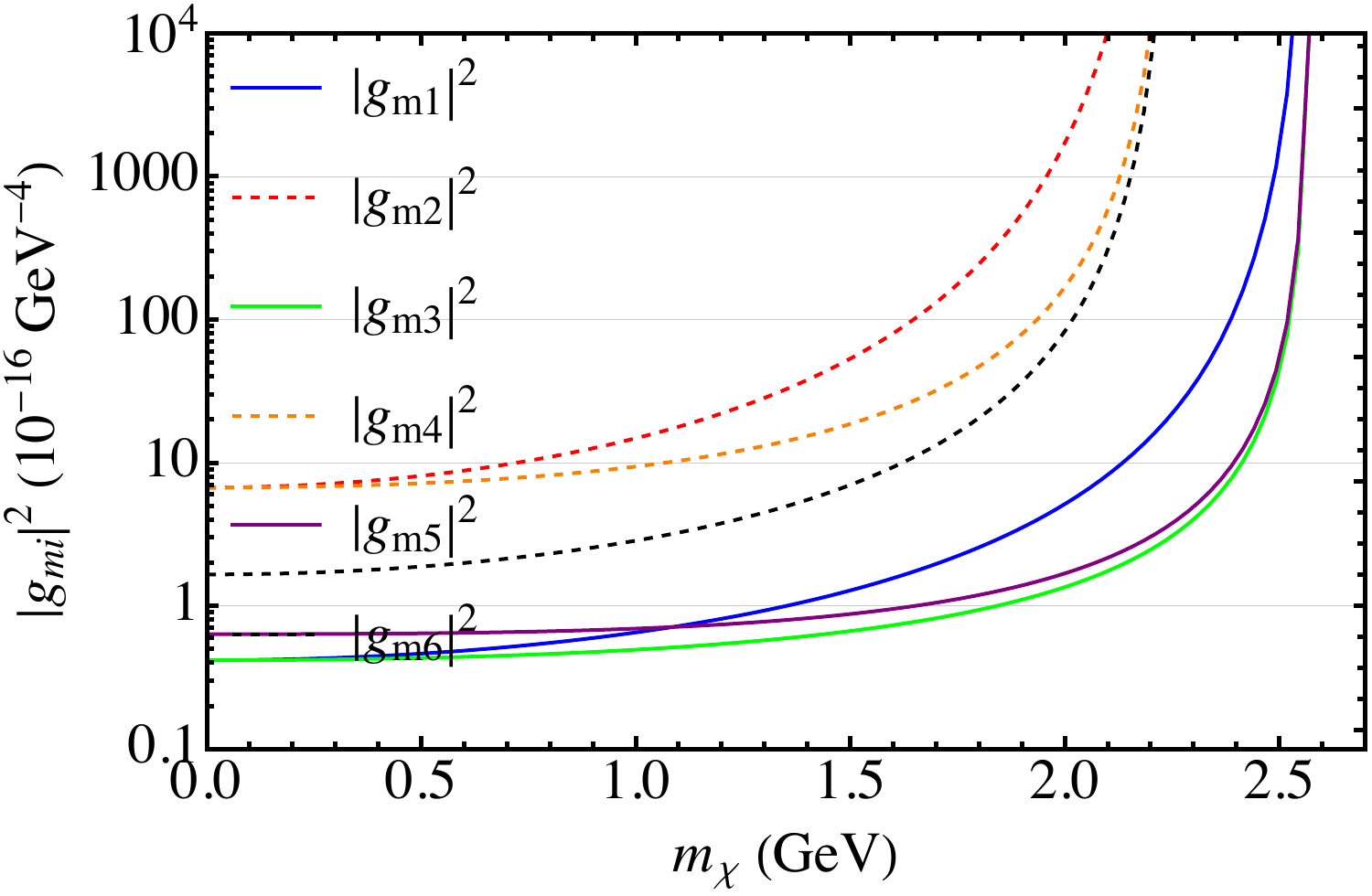}} \\
	\caption{Upper limits of $|g_{mi}|^2$ as functions of $m_\chi$}
	\label{gm-12}
\end{figure}

In our calculation, we assume that only one operator contributes to the process at a time. By integrating the three-body phase space given in Eq.~\eqref{ps3}, the upper limits of the coupling constants $g_{mi}$ can be obtained from Table~{\ref{exp}}, given by
\begin{equation}
	\begin{aligned}
		\mathcal B(M\to M ^{(*)}_f\slashed E)_{\rm exp}  - \mathcal B(M\to M ^{(*)}_f  \bar\nu\nu)_{\rm SM} \ge \mathcal {B}(M\to M ^{(*)}_f\chi\chi)_{Q_i}=\frac{|g_{mi}|^2\widetilde\Gamma_{ii}}{\Gamma_{M_B}},
		\label{eq20}
	\end{aligned}
\end{equation}
where $i=1-6$, the subscript $Q_i$ indicates that this operator contributes at this time, $\widetilde\Gamma_{ii}$ are independent of the coupling constants, and $\Gamma_{M_B}$ is the total width of the initial $B$ meson. Notably, the partial decay width should be divided by two since the Majorana fermion is identical to its antiparticle. The upper limits of $|g_{mi}|^2$ on  ($b s\chi\chi$) and ($bd\chi\chi$) vertices are shown as functions of $m_\chi$ in Fig.~\ref{gm-12} with $m_\chi$ is the mass of $\chi$. 
One can see that when $m_\chi\to 0$, the upper limits of $|g_{mi}|^2$ are $\mathcal O(10^{-17})$ to $\mathcal O(10^{-16})$. Note that the limits of $|g_{m2,4,6}|^2$ are larger than these of $|g_{m1,3,5}|^2$, because the experimental upper bounds on the meson decay processes of $0^-\to1^-$ are larger than those of $0^-\to0^-$ given in Table~\ref{exp}. When $m_\chi$ is larger, the bounds are getting looser as the phase space decreases.

\subsection{Results with invisible particles}

\begin{figure}[h]
	\centering
	\subfigure[~$\Lambda_b\to \Lambda\chi\chi$]{
		\label{width-1}
		\includegraphics[width=0.45\textwidth]{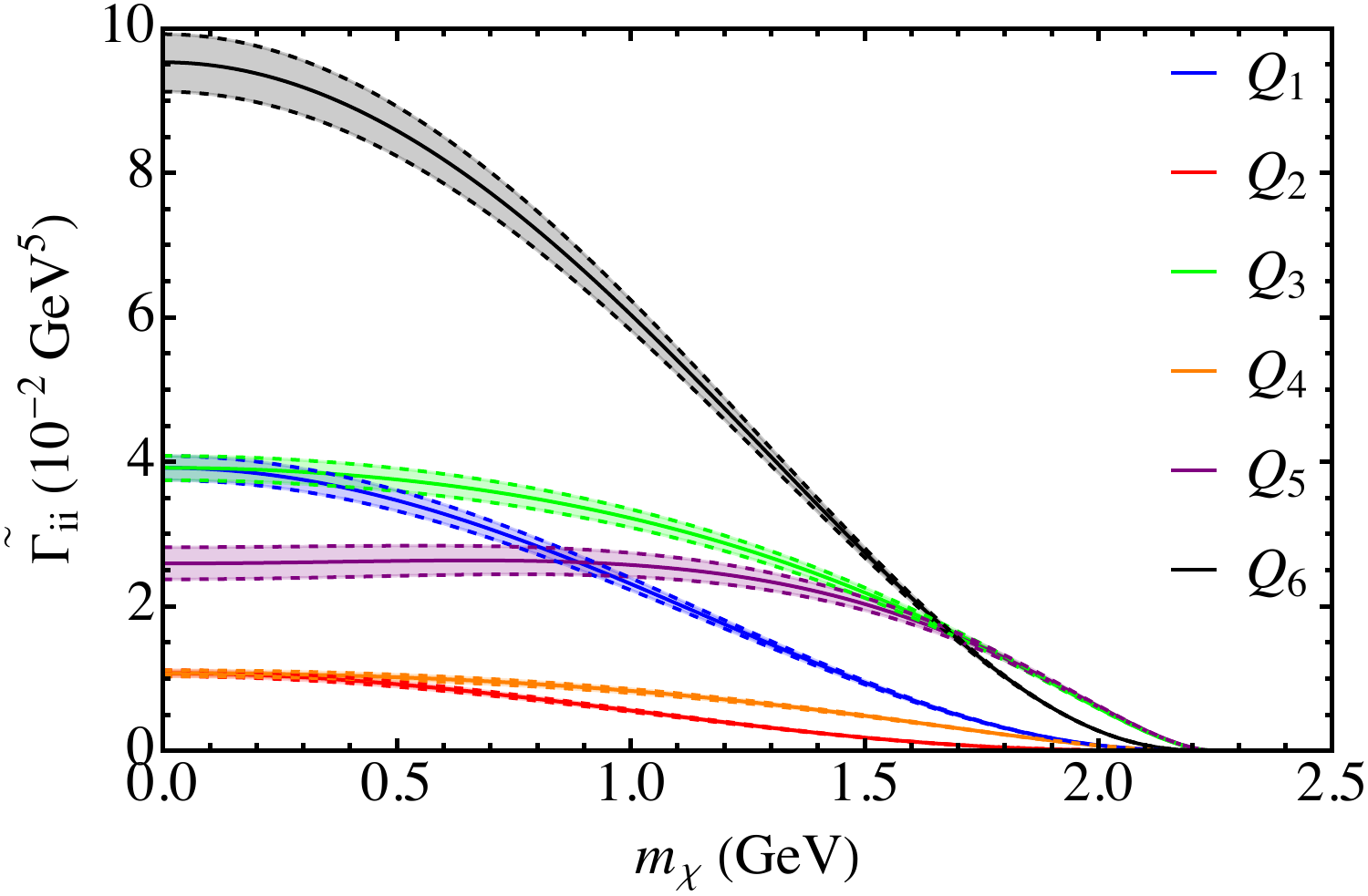}} 
	\hspace{0.5cm}
	\subfigure[~$\Xi_b^{0(-)}\to \Xi^{0(-)}\chi\chi$]{
		\label{width-2}
		\includegraphics[width=0.45\textwidth]{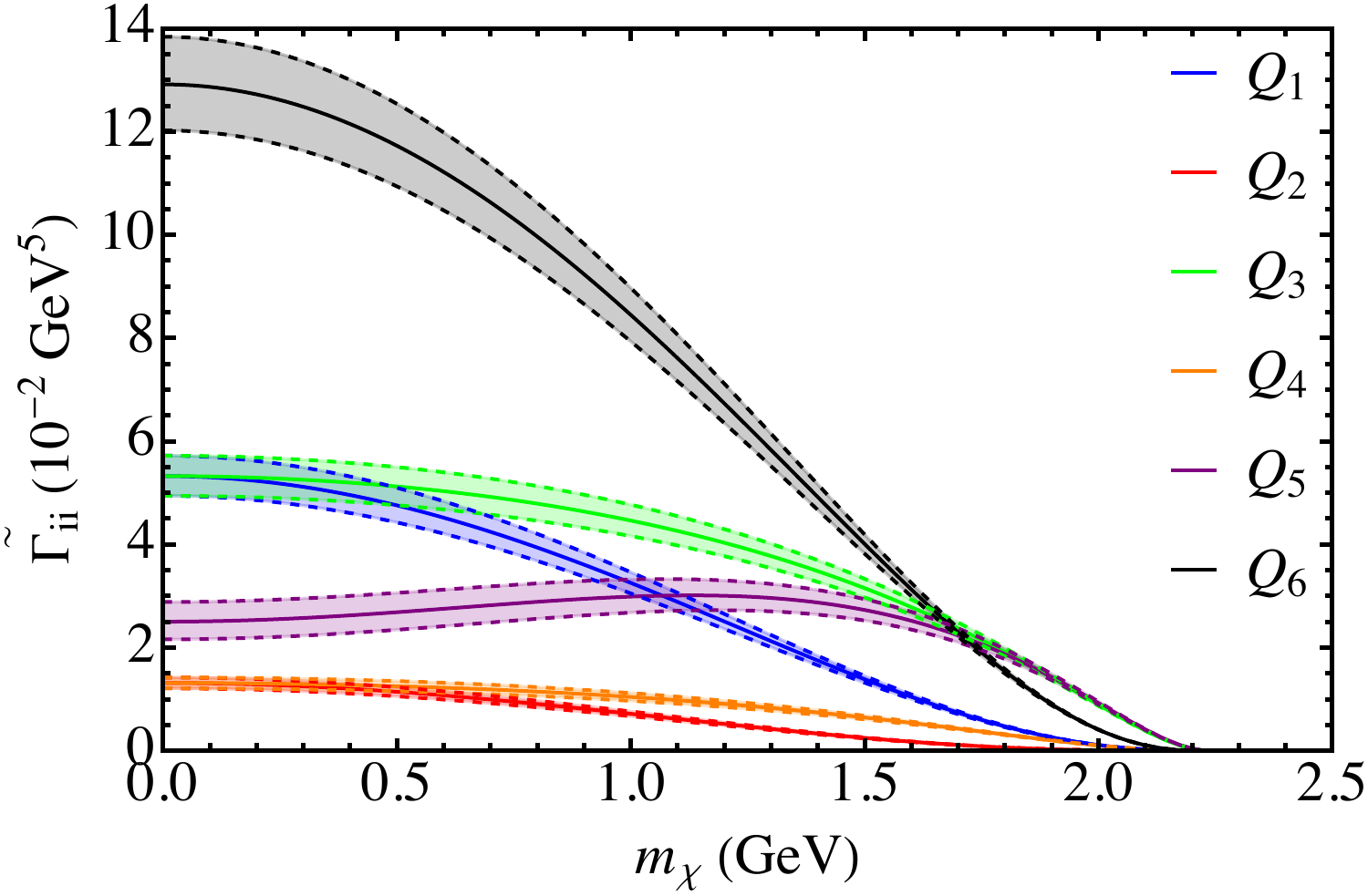}} \\
	\subfigure[~$\Lambda_b\to n\chi\chi$]{
		\label{width-3}
		\includegraphics[width=0.45\textwidth]{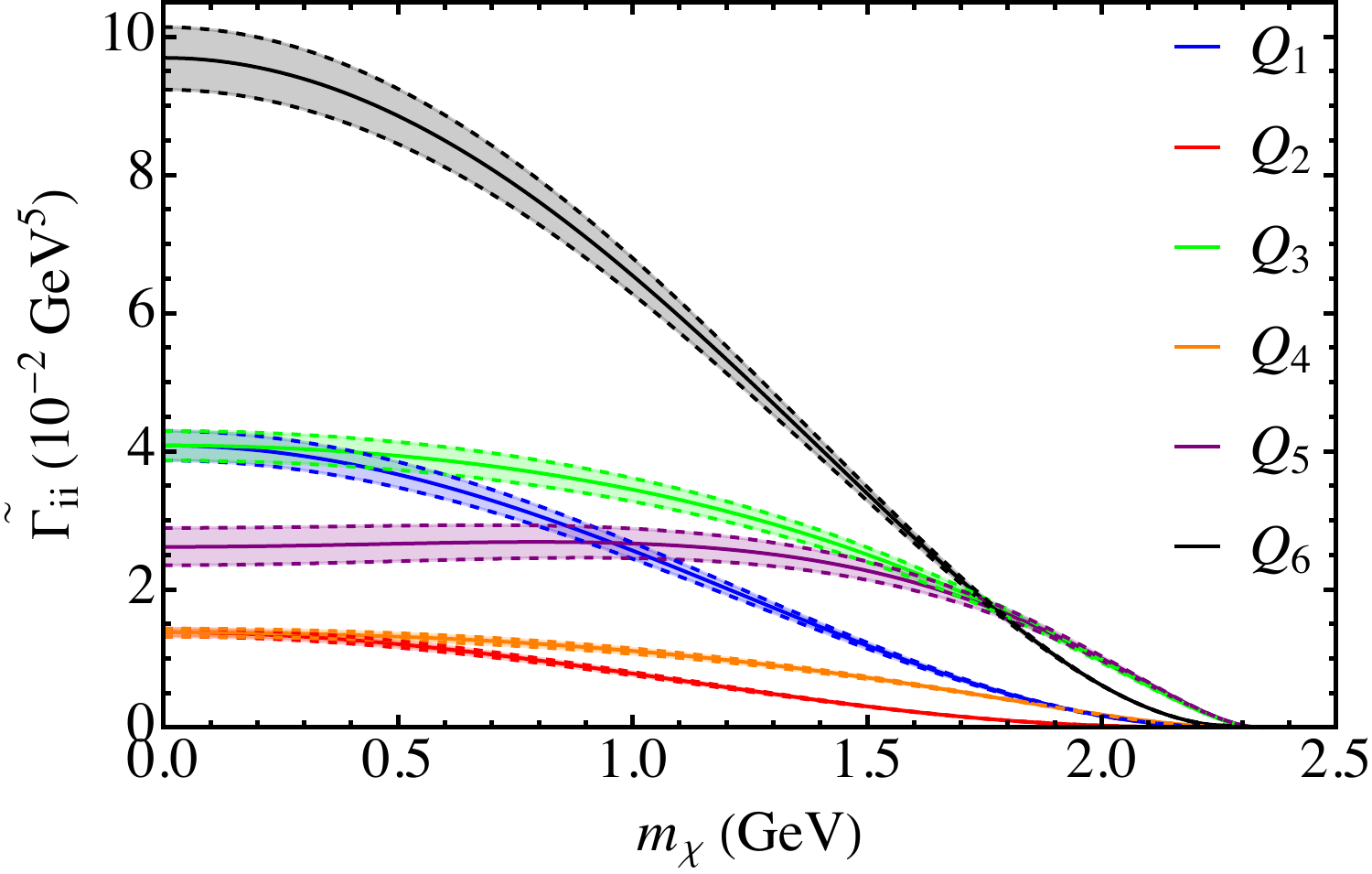}} 
	\hspace{0.5cm}
	\subfigure[~$\Xi_b^-\to \Sigma^-\chi\chi$]{
		\label{width-4}
		\includegraphics[width=0.45\textwidth]{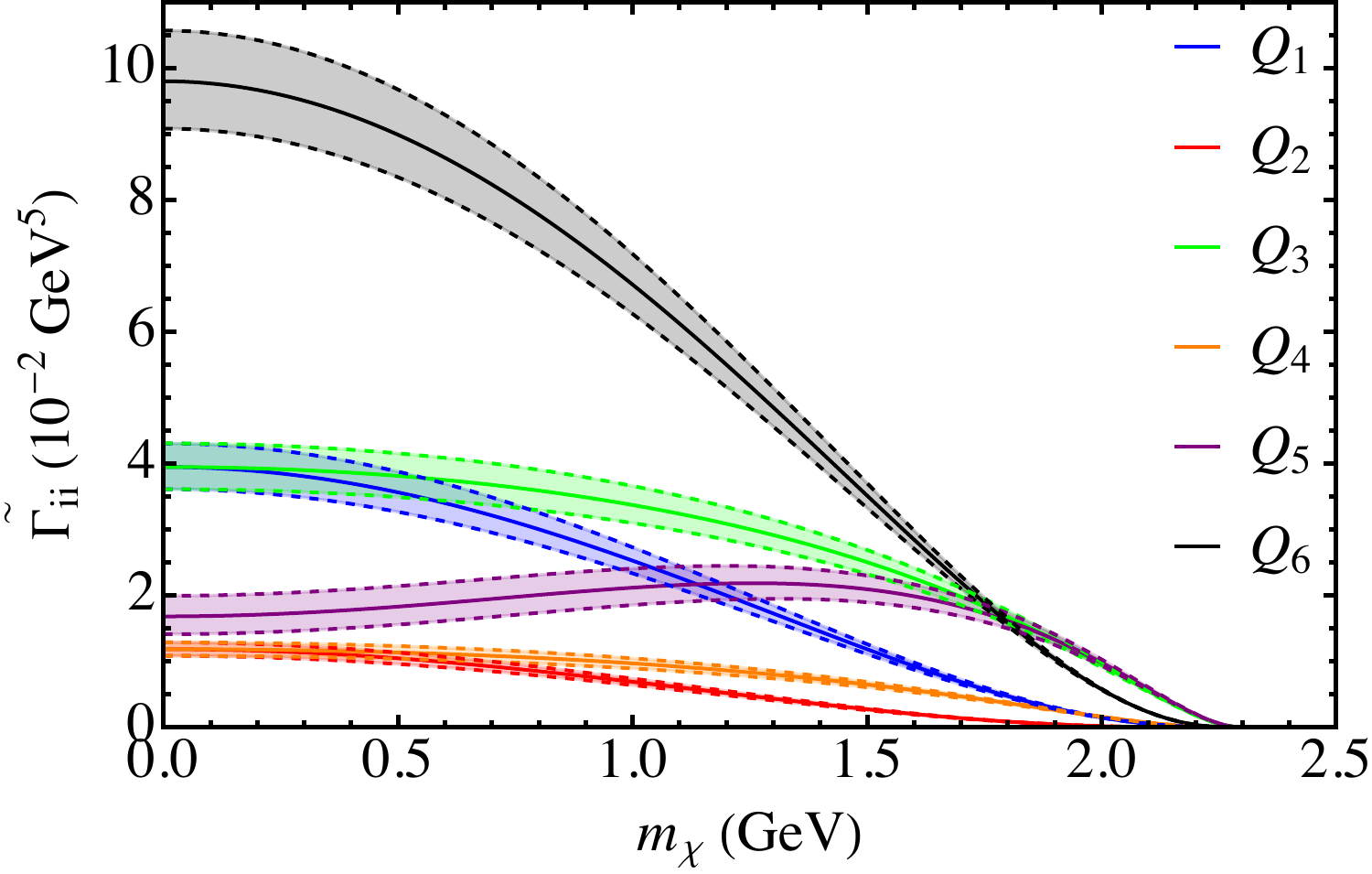}} \\
	\subfigure[~$\Xi_b^0\to \Sigma^0\chi\chi$]{
		\label{width-5}
		\includegraphics[width=0.45\textwidth]{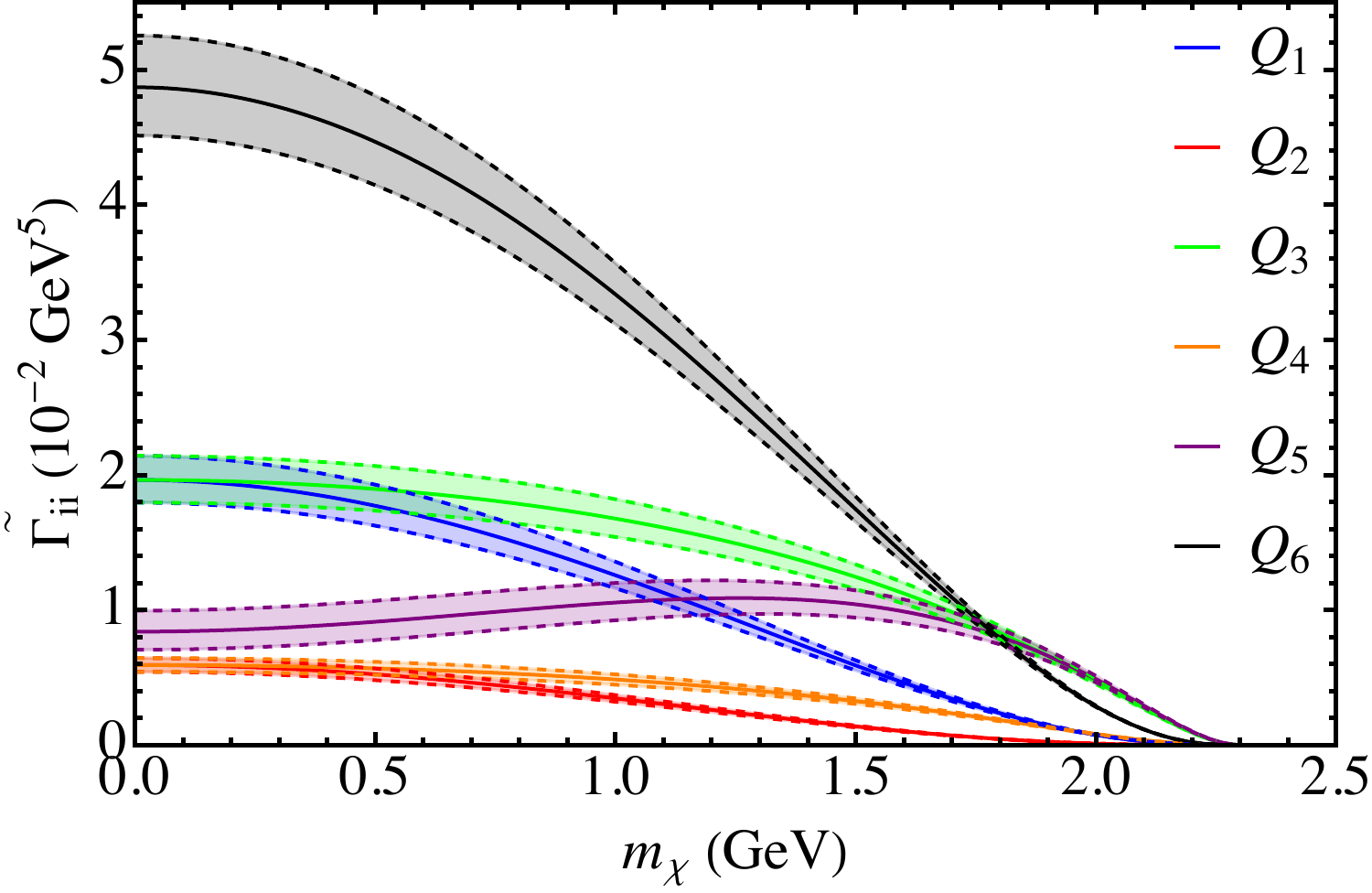}} 
	\hspace{0.5cm}
	\subfigure[~$\Xi_b^0\to \Lambda\chi\chi$]{
		\label{width-6}
		\includegraphics[width=0.45\textwidth]{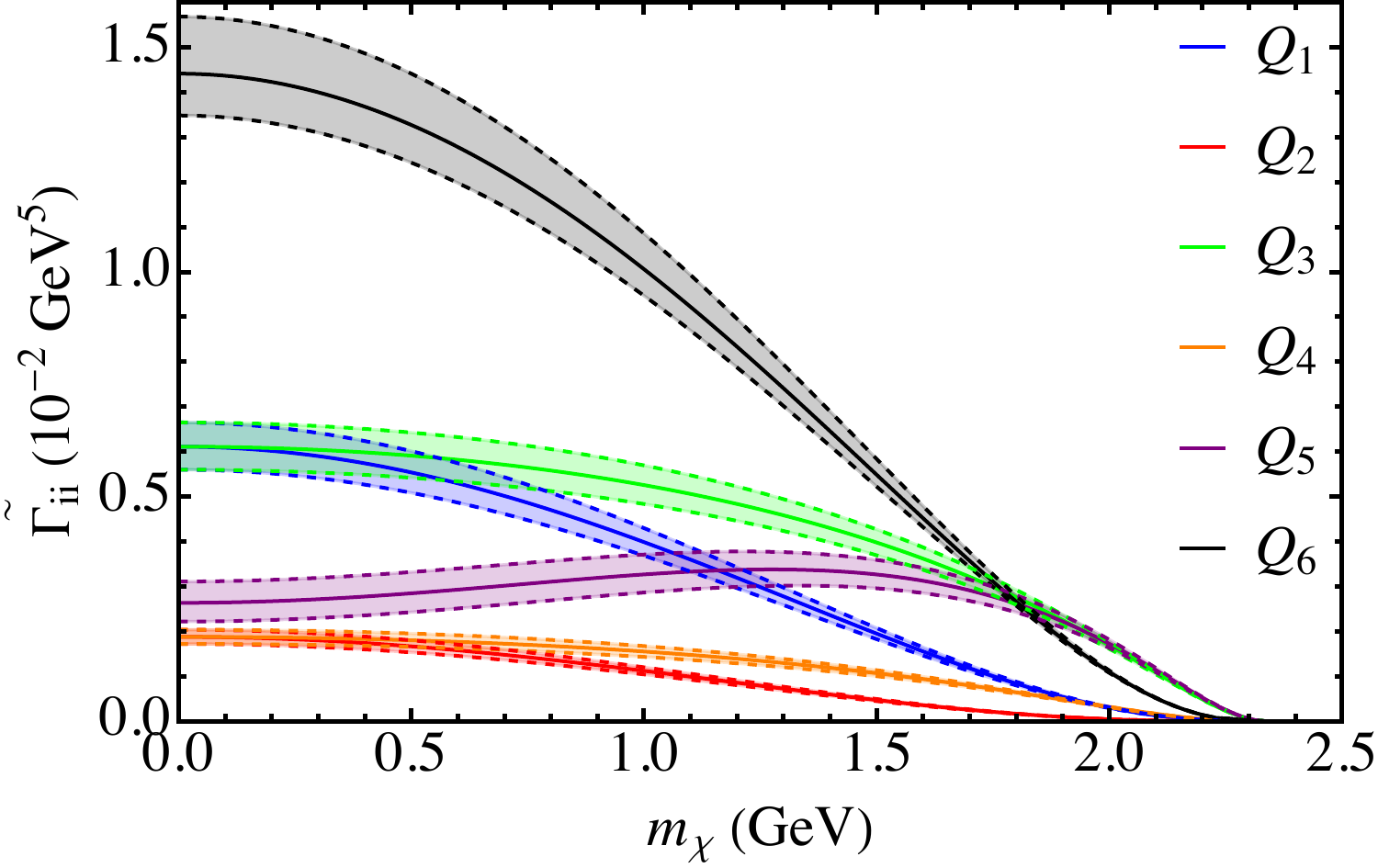}} 
	\caption{$\widetilde\Gamma_{ij}$ as functions of $m_\chi$, where the shadows represent the errors estimated by varying the bag radius within $\pm5\%$}
	\label{width-16}
\end{figure}

For the baryonic decays of $ {\bf B}_b\to {\bf B}_n\chi\chi$, all operators in Eq.~\eqref{op} should be considered. The decay amplitude can be expressed as
\begin{equation}
	\begin{aligned}
		\langle {\bf B}_n\chi\chi|\mathcal L_{eff}| {\bf B}_b\rangle
		&=2g_{m1}  \langle {\bf B}_n|(\bar q_{_f} q) |{\bf B}_b\rangle\bar u_{\chi}v_{\chi}+2g_{m2} \langle {\bf B}_n|(\bar q_{_f}\gamma^5 q) |{\bf B}_b\rangle\bar u_{\chi}v_{\chi}\\
		&+2 g_{m3} \langle {\bf B}_n|(\bar q_{_f} q) |{\bf B}_b\rangle\bar u_{\chi}\gamma^5 v_{\chi}
		+2g_{m4}  \langle {\bf B}_n|(\bar q_{_f}\gamma^5 q) |{\bf B}_b\rangle\bar u_{\chi}\gamma^5 v_{\chi}\\
		&+2g_{m5} \langle {\bf B}_n|(\bar q_{_f}\gamma_\mu q) |{\bf B}_b\rangle\bar u_{\chi}\gamma^{\mu}\gamma^5 v_{\chi}
		+2g_{m6} \langle {\bf B}_n|(\bar q_{_f}\gamma_\mu \gamma^5q) |{\bf B}_b\rangle\bar u_{\chi}\gamma^{\mu}\gamma^5 v_{\chi}.   \\
	\end{aligned}
\end{equation}
Here, the baryonic transition matrix elements have been given by Eq.~\eqref{ffb}, while the numerical values of the FFs have been shown in Figs.~\ref{ff-1}-\ref{ff-6}. As above, we discuss the contribution of each operator separately. By integrating the three-body phase space in Eq.~\eqref{ps3}, $\widetilde\Gamma_{ii}$ defined in Eq.~\eqref{eq20} are obtained with the numerical results in Fig.~\ref{width-16}. 
One can see that $\widetilde\Gamma_{11,22,33,44,66}$ decrease to zero as $m_\chi$ increases due to the phase space reduction, while $\widetilde\Gamma_{55}$ increases first and then decreases to zero. The upper bound of $m_\chi$ can be taken as $(M-M_f)/2$. When $m_\chi=0$, we have that $\widetilde\Gamma_{11}=\widetilde\Gamma_{33}$ and $\widetilde\Gamma_{22}=\widetilde\Gamma_{44}$, since $\widetilde\Gamma_{11,22}$ and $\widetilde\Gamma_{33,44}$ are proportional to $(P_1\cdot P_2-m_\chi^2)$ and $(P_1\cdot P_2+m_\chi^2)$, respectively. The uncertainties of $\widetilde\Gamma_{ii}$ are about $\pm10\%$. It should be noted that $\widetilde\Gamma_{ii}$ are independent of the coupling constants. 

\begin{figure}[h]
	\centering
	\subfigure[~$\Lambda_b\to \Lambda\chi\chi$]{
		\label{br-1}
		\includegraphics[width=0.45\textwidth]{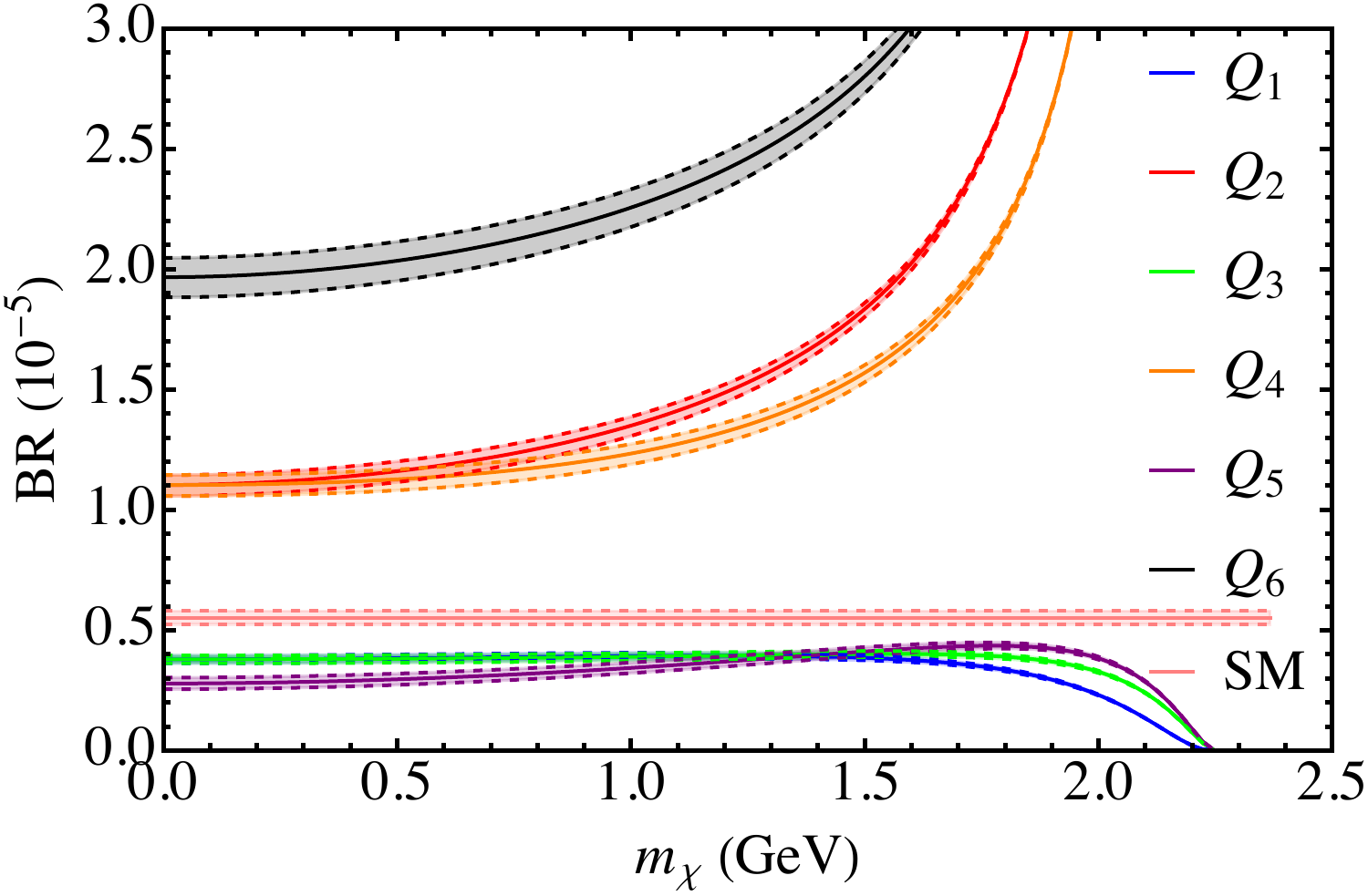}} 
	\hspace{0.5cm}
	\subfigure[~$\Xi_b^{0(-)}\to \Xi^{0(-)}\chi\chi$]{
		\label{br-2}
		\includegraphics[width=0.45\textwidth]{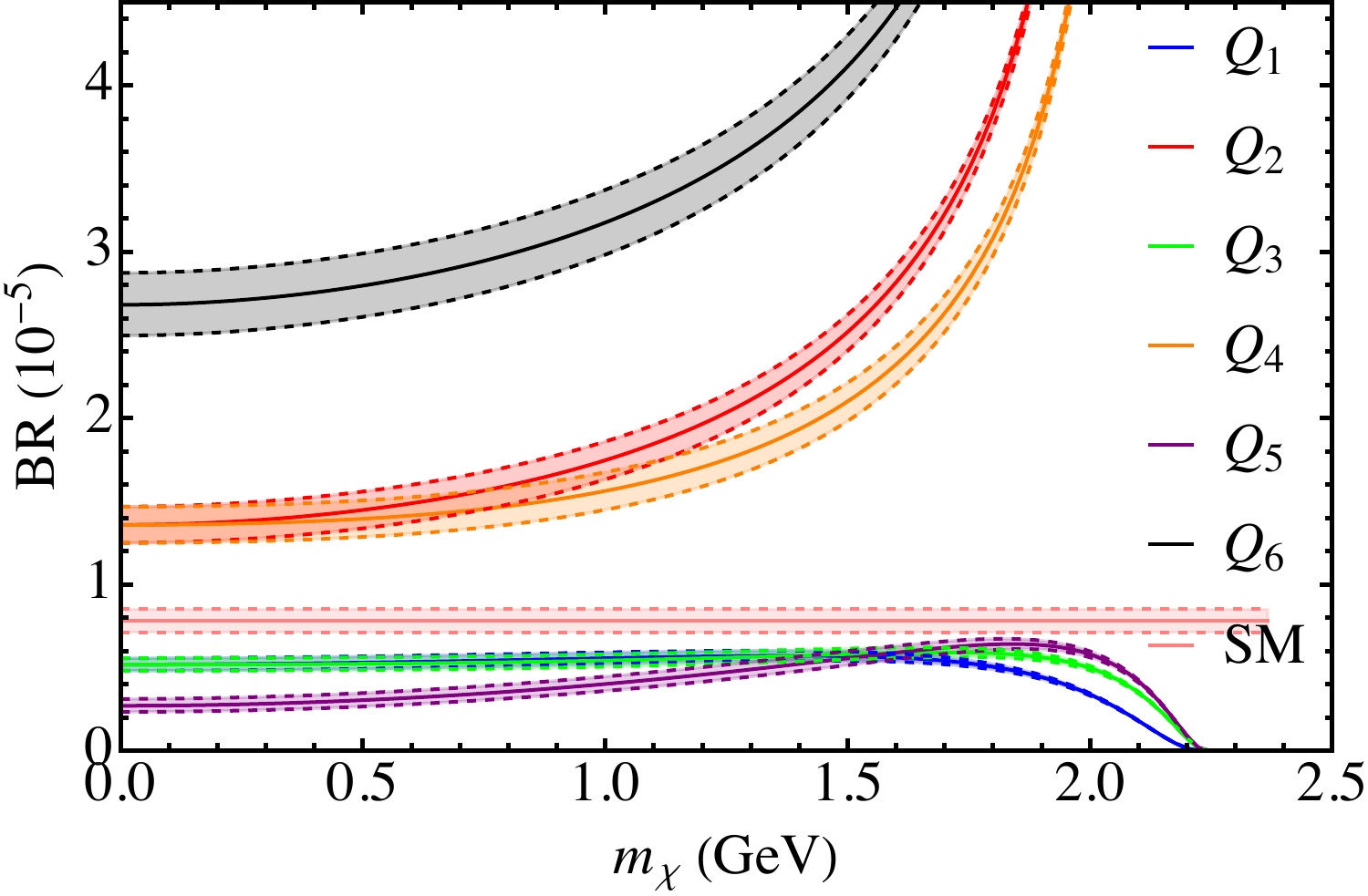}} \\
	\subfigure[~$\Lambda_b\to n\chi\chi$]{
		\label{br-3}
		\includegraphics[width=0.45\textwidth]{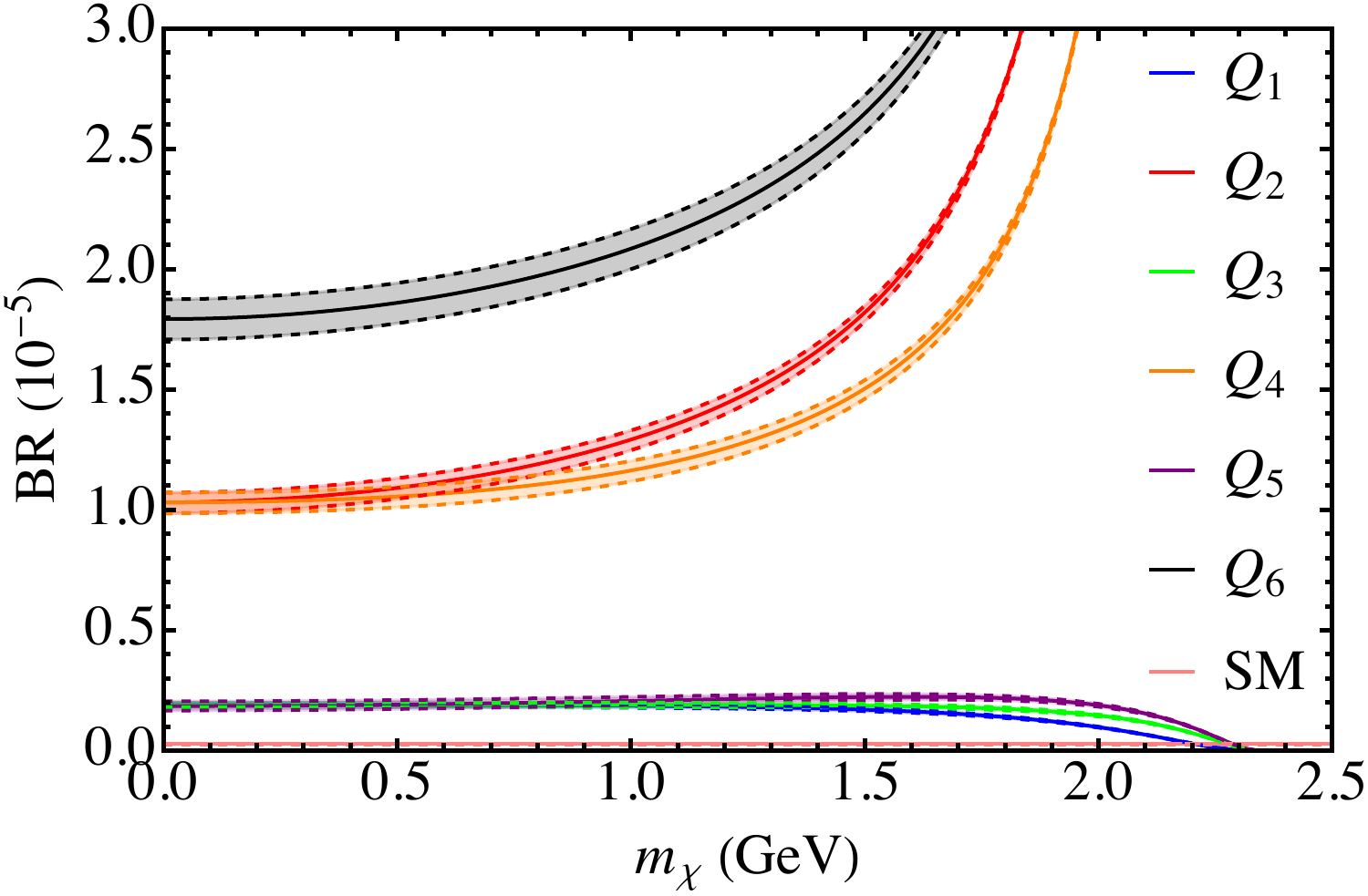}} 
	\hspace{0.5cm}
	\subfigure[~$\Xi_b^-\to \Sigma^-\chi\chi$]{
		\label{br-4}
		\includegraphics[width=0.45\textwidth]{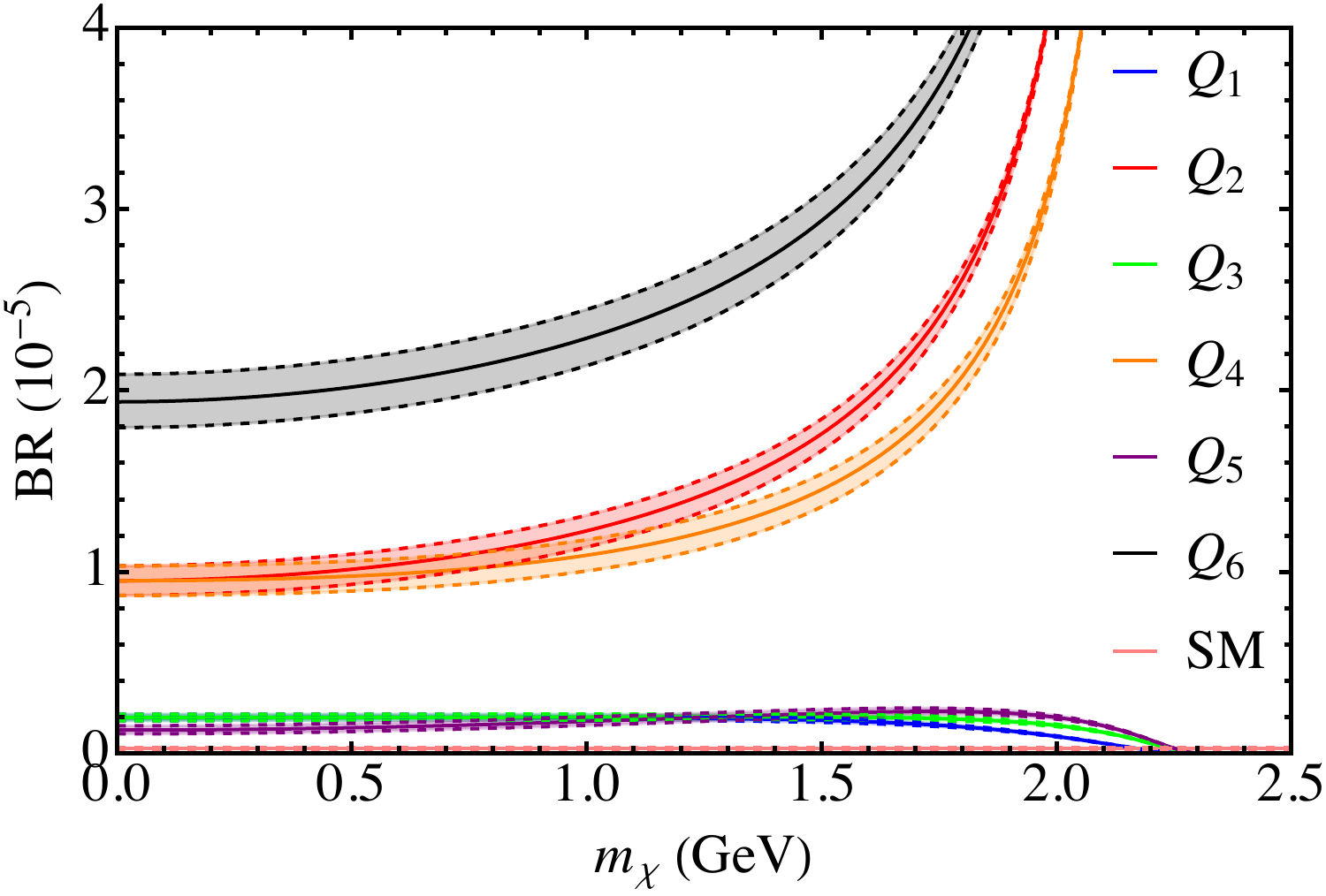}} \\
		\subfigure[~$\Xi_b^0\to \Sigma^0 \chi\chi$]{
		\label{br-5}
	\includegraphics[width=0.45\textwidth]{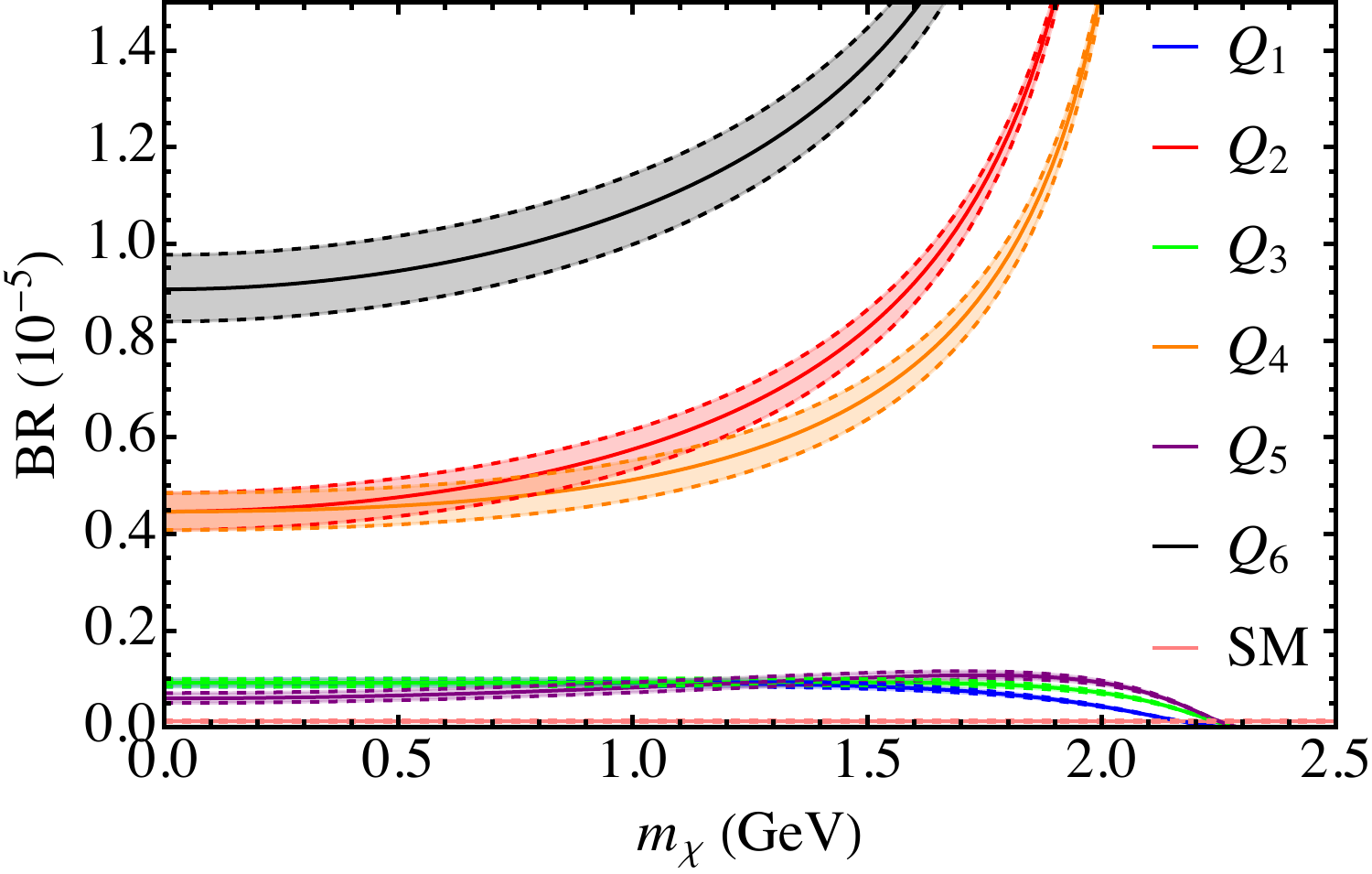}} 
		\hspace{0.5cm}
	\subfigure[~$\Xi_b^0\to \Lambda\chi\chi$]{
		\label{br-6}
		\includegraphics[width=0.45\textwidth]{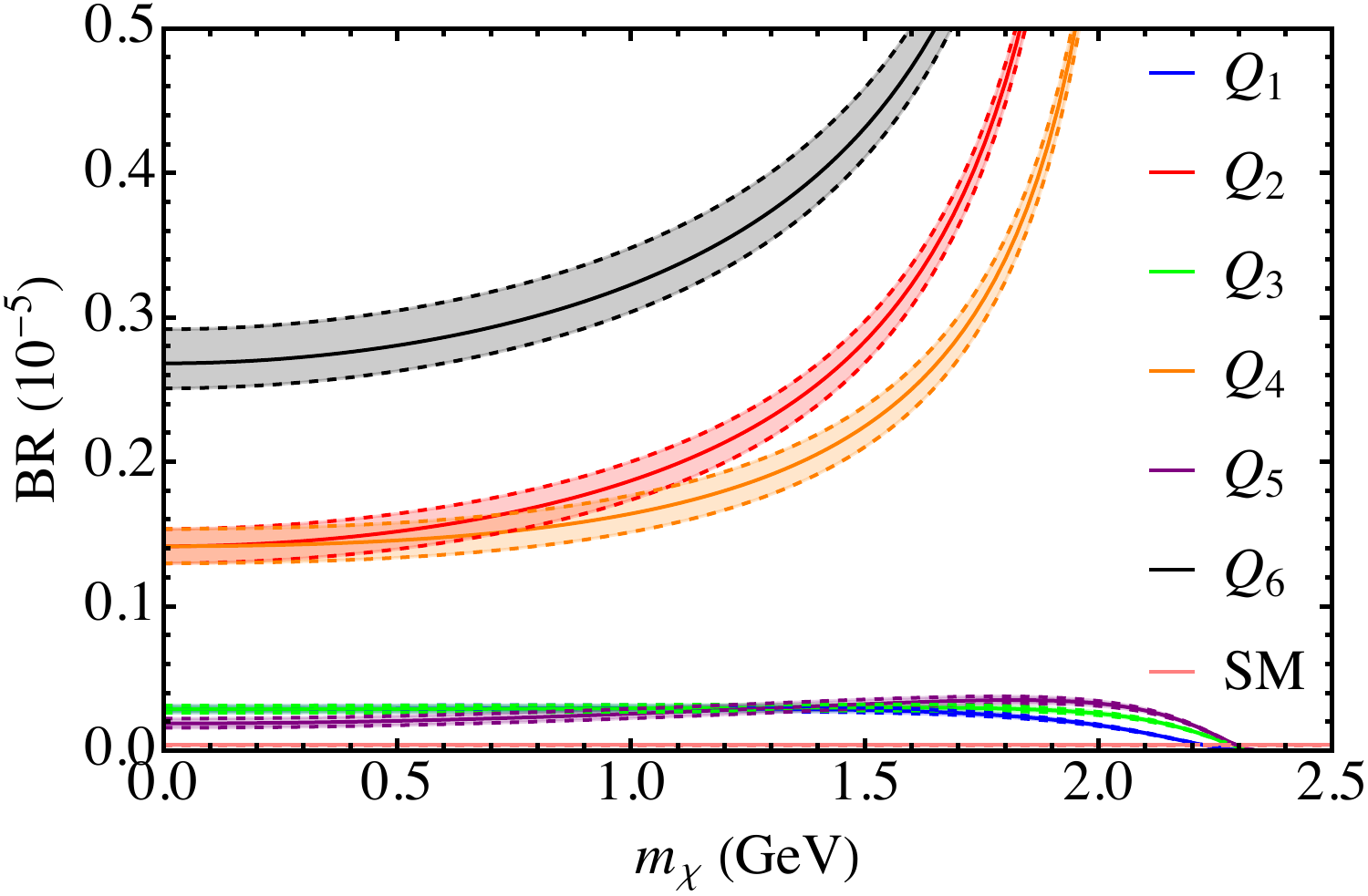}} 
	\caption{Upper limits of $\mathcal {B}({\bf B}_b\to{\bf B}_n\chi\chi)$ as functions of $m_\chi$, where the shadows represent the errors estimated by varying the bag radius within $\pm5\%$}
	\label{br-16}
\end{figure}

By combining $\widetilde\Gamma_{ii}$ with the bounds of the coupling coefficients given in Fig.~\ref{gm-12}, we obtain the upper limits of the decay branching ratios associated with the SM predictions, as shown in Fig.~\ref{br-16}. 
We see that in the most regions of $m_\chi$, the upper limits of the branching ratios are $\mathcal O(10^{-6})$ to $\mathcal O(10^{-5})$, which are about of the same orders or an order of magnitude larger than the SM expectations of $10^{-8}$ to $10^{-6}$. In Figs.~\ref{br-16}c-\ref{br-16}f, the solid pink lines representing the SM are close to the $X$ axis. In particular, $Q_{2,4,6}$ make the dominant contributions. This is because the bounds on $|g_{m2,4,6}|^2$ are looser than these on $|g_{m1,3,5}|^2$. When $m_\chi\to(M-M_f)/2$, the upper limits for the branching ratios from $Q_{2,4,6}$ approach infinity, because the mass difference between the initial and final mesons is smaller than that between the initial and final baryons. For a larger value of $m_\chi$, the baryon decays cannot be limited by the meson decay channels. 

In Tables~\ref{br0}, \ref{br1} and \ref{br2}, we list the central values of upper limits of $\mathcal B({\bf B}_b\to{\bf B}_n\chi\chi)$ for $m_\chi=0,~1$ and $2$ GeV, respectively. In Table~\ref{br0}, we also show the SM predictions of $\mathcal B({\bf B} _b\to{\bf B}_n\bar\nu\nu)$. 
\begin{table}[h]
	\setlength{\tabcolsep}{0.1cm}
	\caption{Upper limits of $\mathcal B({\bf B} _b\to{\bf B}_n\chi\chi)$ when $m_\chi\to 0~{\rm GeV}$ (in units of $10^{-5}$)}
	\centering
	\begin{tabular*}{\textwidth}{@{}@{\extracolsep{\fill}}ccccccc}
		\hline\hline
		Operator&$\Lambda_b\to \Lambda\chi\chi$&$\Xi_b^{0(-)}\to \Xi^{0(-)}\chi\chi$&$\Lambda_b\to n\chi\chi$&$\Xi_b^-\to \Sigma^-\chi\chi$&$\Xi_b^0\to \Sigma^0\chi\chi$&$\Xi_b^0\to \Lambda\chi\chi$\\
		\hline
		$Q_1$& $0.38$  & $0.52$ &$0.19$ &$0.20$ &$0.092$& $0.029$  \\
		$Q_2$& $1.1$  & $1.4$ &$1.0$ &$0.95$ &$0.45$& $0.14$  \\
		$Q_3$& $0.38$  & $0.52$ &$0.19$ &$0.20$ &$0.092$& $0.029$  \\
		$Q_4$& $1.1$  & $1.4$ &$1.0$ &$0.95$ &$0.45$ & $0.14$ \\
		$Q_5$ &$0.28$ &$0.27$ & $0.19$ &$0.13$ &$0.060$&$0.019$ \\
		$Q_6$& $2.0$  & $2.7$ &$1.8$ &$1.9$ &$0.91$& $0.27$  \\
		\hline
		SM&$\Lambda_b\to \Lambda\bar\nu\nu$&$\Xi_b^{0(-)}\to \Xi^{0(-)}\bar\nu\nu$&$\Lambda_b\to n\bar\nu\nu$&$\Xi_b^-\to \Sigma^-\bar\nu\nu$&$\Xi_b^0\to \Sigma^0\bar\nu\nu$&$\Xi_b^0\to \Lambda\bar\nu\nu$\\
		& $0.55$  & $0.78$ &$0.028$ &$0.027$ &$0.012$ & $0.0039$ \\
		\hline\hline
		\label{br0}
	\end{tabular*}
\end{table}
\begin{table}[h]
	\setlength{\tabcolsep}{0.1cm}
	\caption{Upper limits of $\mathcal B({\bf B} _b\to{\bf B}_n\chi\chi)$ when $m_\chi= 1~{\rm GeV}$ (in units of $10^{-5}$)}
	\centering
	\begin{tabular*}{\textwidth}{@{}@{\extracolsep{\fill}}ccccccc}
		\hline\hline
		Operator&$\Lambda_b\to \Lambda\chi\chi$&$\Xi_b^{0(-)}\to \Xi^{0(-)}\chi\chi$&$\Lambda_b\to n\chi\chi$&$\Xi_b^-\to \Sigma^-\chi\chi$&$\Xi_b^0\to \Sigma^0\chi\chi$&$\Xi_b^0\to \Lambda\chi\chi$\\
		\hline
		$Q_1$& $0.39$  & $0.56$ &$0.19$ &$0.20$ &$0.092$ & $0.029$ \\
		$Q_2$& $1.4$  & $1.8$ &$1.3$ &$1.2$ &$0.58$& $0.19$  \\
		$Q_3$& $0.39$  & $0.54$ &$0.19$ &$0.20$ &$0.093$& $0.029$  \\
		$Q_4$& $1.2$  & $1.6$ &$1.2$ &$1.1$ &$0.51$& $0.16$  \\
		$Q_5$ &$0.34$ &$0.40$ & $0.21$ &$0.17$ &$0.082$&$0.025$ \\
		$Q_6$& $2.3$  & $3.2$ &$2.1$ &$2.3$  &$1.1$& $0.32$ \\
		\hline\hline
		\label{br1}
	\end{tabular*}
\end{table}
\begin{table}[h]
	\setlength{\tabcolsep}{0.1cm}
	\caption{Upper limits of $\mathcal B({\bf B} _b\to{\bf B}_n\chi\chi)$ when $m_\chi= 2~{\rm GeV}$ (in units of $10^{-5}$)}
	\centering
	\begin{tabular*}{\textwidth}{@{}@{\extracolsep{\fill}}ccccccc}
		\hline\hline
		Operator&$\Lambda_b\to \Lambda\chi\chi$&$\Xi_b^{0(-)}\to \Xi^{0(-)}\chi\chi$&$\Lambda_b\to n\chi\chi$&$\Xi_b^-\to \Sigma^-\chi\chi$&$\Xi_b^0\to \Sigma^0\chi\chi$&$\Xi_b^0\to \Lambda\chi\chi$\\
		\hline
		$Q_1$& $0.22$  & $0.33$ &$0.096$ &$0.091$ &$0.042$& $0.017$  \\
		$Q_2$& $5.3$  & $7.3$ &$5.2$ &$4.4$ &$2.0$& $0.93$  \\
		$Q_3$& $0.32$  & $0.49$ &$0.14$ &$0.15$ &$0.071$& $0.026$  \\
		$Q_4$& $3.6$  & $5.4$ &$3.6$ &$3.3$ &$1.6$& $0.61$  \\
		$Q_5$ &$0.38$ &$0.57$ & $0.19$ &$0.20$&$0.091$&$0.032$ \\
		$Q_6$& $6.3$  & $9.2$ &$5.7$ &$5.8$ &$2.7$& $1.0$  \\
		\hline\hline
		\label{br2}
	\end{tabular*}
\end{table}
We find that for the decays with $b\to s$ transition, the contributions from the new operators are almost of the same orders as the SM ones. While for these with $b\to d$ transition, the upper bounds of the decay modes with the invisible particles are about one to two orders of magnitude larger than $\mathcal {B}({\bf B}_b\to{\bf B}_n\bar\nu\nu)$ due to the CKM matrix element depressions. Clearly, it is more hopeful to distinguish new neutral particles from the SM neutrinos experimentally. When $m_\chi$ is larger, the upper limits of the contributions from $Q_{2,4,6}$ are getting looser. The upper limits of the branching ratios of decay modes with invisible particles are estimated to be $\mathcal O(10^{-5})-\mathcal O(10^{-6})$. We expect that in the near future, experiments on the bottomed baryon FCNC decays could give more relevant results for comparisons. 

%%%%%%%%%%%%%%%%%%%%%%%%%%%%%%%%%%
\section{Conclusion}
%%%%%%%%%%%%%%%%%%%%%%%%%%%%%%%%%%
We have studied the light invisible Majorana fermions in the FCNC processes of the long-lived bottomed baryons. The model-independent effective Lagrangian which contains six operators has been introduced to describe the couplings between the quarks and invisible Majorana fermions. The bounds of the coupling constants have been extracted from the differences between the experimental upper limits and SM predictions of the relevant $B$ meson FCNC decays. Based on these bounds, we have predicted the upper limits of $\mathcal B({\bf B}_b\to{\bf B}_n\chi\chi)$. In particular, we have found that the decay branching ratios of $\Lambda_b\to\Lambda\chi\chi$, $\Xi_b^{0(-)}\to\Xi^{0(-)}\chi\chi$, $\Lambda_b\to n \chi\chi$, $\Xi_b^{-}\to \Sigma^{-} \chi\chi$, $\Xi_b^{0}\to \Sigma^{0} \chi\chi$, and $\Xi_b^{0}\to \Lambda \chi\chi$ can be as large as  $(2.0,~2.7,~1.8,~1.9,~0.91,~0.27)\times10^{-5}$, $(2.3,~3.2,~2.1,~2.3,~1.1,~0.32)\times10^{-5}$, and $(6.3,~9.2,~5.7,~5.8,~2.7,~1.0)\times10^{-5}$ with $m_\chi=0$, $1$, and $2$ GeV, respectively. We are looking forward to the future experiments, such as those at Belle II, to get more measurements on bottomed baryons to find signs of new particles.

%%%%%%%%%%%%%%%%%%%%%%%%%%%%%%%%%%%%
\section{Acknowledgments}
%%%%%%%%%%%%%%%%%%%%%%%%%%%%%%%%%%%%
This work is supported in part by the National Key Research and Development Program of China under Grant No.~2020YFC2201501 and the National Natural Science Foundation of China (NSFC) under Grant No.~12147103.

\bibliography{reference}

\end{document}